\documentstyle[12pt,leqno]{article}

\newfont{\bb}{msbm10 scaled\magstep1}
\newfont{\frak}{eufm10 scaled\magstep1}

\newcommand{\noi}{\noindent}

\newcommand{\ket}[1]{|#1\rangle}
\newcommand{\braket}[2]{\langle #1|#2\rangle}

\newcommand{\be}{\begin{equation}}
\newcommand{\ee}{\end{equation}}
\newcommand{\h}{\hbar}

\newcommand{\bbC}{\mbox{\bb C}}
\newcommand{\bbR}{\mbox{\bb R}}

\newcommand{\bbZ}{\mbox{\bb Z}}

\newcommand{\frakp}{\mbox{\frak p}}

\newcommand{\frakt}{\mbox{\frak t}}

\newcommand{\lig}{\mbox{\frak g}}

\newcommand{\tr}{\mbox{Tr}}

\newcommand{\tildx}{\skew4\tilde x}

\newcommand{\por}{{\cdot}}

\newcommand{\SL}{\mbox{SL}}

\newcommand{\calP}{{\cal P}}

\newcommand{\calB}{{\cal B}}
\newcommand{\calF}{{\cal F}}
\newcommand{\calG}{{\cal G}}
\newcommand{\calS}{{\cal S}}

\newcommand{\calH}{{\cal H}}

\newcommand{\calV}{{\cal V}}
\newcommand{\calC}{{\cal C}}

\newcommand{\calM}{{\cal M}}
\newcommand{\calK}{{\cal K}}

\newtheorem{theorem}{Theorem}[section]
\newtheorem{corollary}[theorem]{Corollary}
\newtheorem{lemma}[theorem]{Lemma}
\newtheorem{proposition}[theorem]{Proposition}

\newtheorem{definition}[theorem]{Definition}

\newenvironment{proof}{ {\sc Proof} \hskip
.2truein \rm}{\nobreak\par\hfill
                  $\Box$
         \vskip .1truein  \par}

\def\today{\ifcase\month\or
  January\or February\or March\or April\or May\or June\or
  July\or August\or September\or October\or November\or December\fi
 \space\number\day, \number\year}

\newcommand{\calZ}{{\cal Z}}

\newcommand{\ol}{\overline}

\title{Legendrian Distributions with Applications to the
Non-Vanishing of Poincar\'e Series of Large Weight}
\author{D. Borthwick\thanks{Mathematics Department,
University of Michigan, Ann Arbor, Michigan 48109 }\
\ T. Paul\thanks {CEREMADE, Universit\'e Paris-Dauphine,
Place de Lattre de Tassigny, 75775 Paris Cedex 16 et CNRS}
\ and A. Uribe\thanks{ Mathematics Department,
University of Michigan, Ann Arbor, Michigan 48109.
Research supported in part by NSF grant DMS-9303778.}}

\begin{document}
\setcounter{page}{0}
\setcounter{footnote}{2}
\date{}
\maketitle
\tableofcontents
\vfill\break
\newcommand{\dL}{\delta_{\Lambda}}
\newcommand{\ip}[1]{\langle #1\rangle}
\section{Introduction}

Let $X$ be a compact K\"ahler manifold of complex dimension $n$
and
K\"ahler form $\Omega$, equipped with a holomorphic Hermitian
line bundle
$L\to X$ such that the curvature of its natural connection is
$\Omega$.  We will refer to such an $L$ as a quantizing line bundle.
For each positive integer $k$, let
\be\label{1.1}
\calS_k\,=\, H^0(X, L^{\otimes k})
\ee
be the complex inner-product space of holomorphic sections of the $k$-
th
tensor power of $L$.  Philosophically, $\calS_k$ is the quantum
phase space of $X$ where $k$ is the inverse of Planck's constant.
In this paper we do the following:
\begin{enumerate}
\item We associate, to certain immersed
Lagrangian submanifolds $\Lambda\to
X$, sequences of sections $u_k\in \calS_k$, $k=1,2,\ldots$.
These sections represent quantum-mechanical states that
are associated semi-classically with $\Lambda$.
The $\Lambda$'s in question (defined below) will be called
{\em Bohr-Sommerfeld Lagrangians}.
\item To each such sequence, we associate a symbol which is a
half-form on $\Lambda$.
\item We compute the large $k$ asymptotics
of matrix coefficients $\ip{Tu_k,u_k}$ where $T$ is a
Toeplitz operator.  The dependence on $T$ of the leading order term is
an integral of the symbol of $T$ over $\Lambda$, proving that these
sections concentrate on $\Lambda$.
By taking $T=I$ we obtain estimates on the $L^2$ norms of the $u_k$.
In particular we show that these sections are not zero for $k$ large.
We also estimate $\ip{Tu_k,v_k}$ where $\{v_k\}$ is a sequence
associated with a second immersed Lagrangian intersecting $\Lambda$
cleanly.
\item  For $X$ a Riemann surface, the elements of $\calS_k$ are
holomorphic cusp forms of weight $2k$.  We show that the
Poincar\'e series associated to geodesics are particular cases of our
construction.  The classical Poincar\'e series associated with cusps are
also
a particular case of our construction (although the analysis
applies only in the manifold case).   As a corollary of the asymptotic
expansion,
we find that the classical Poincar\'e series of large
weight associated to a fixed periodic geodesic is non-vanishing for large
weight.  We extend this result to hypercycles and circles.
\end{enumerate}

\noi
Our main results are Theorem \ref{mainthm}, which gives the
asymptotics of the matrix elements, and Theorem \ref{localuk}, in
which we establish  the local realization:
\be
u_k (x) = k^{n/2} (\hbox{\it Gaussian}) + O(k^{(n-1)/2}).
\ee

\medskip
We thus establish a precise correspondence between Bohr-Sommerfeld
Lagrangian submanifolds of $X$ (equipped with half-forms, see below)
and certain sequences of states depending on $k=1/\h$.  Motivation for
this
comes from general quantization/semi-classical ideas in the context
of Kahler phase spaces.  As were many others working in this area, we
were very influenced by the pioneering work of F. A. Berezin, \cite{B}.
He was one of the first to study the semi-classical (i.e. large $k$) limit
of Toeplitz operators with multipliers given by functions on $X$.
For further developments of Berezin's ideas, see \cite{CGR} and references
therein, and also \cite{KL}, \cite{BLU}, \cite{BMS}.
To our knowledge, no systematic method of quantization of
Bohr-Sommerfeld Lagrangians has been developed.  In addition to the
applications to Poincar\'e series presented in \S 4 of this
paper, our construction can be applied, e.g., to the quantization and
semi-classical limit of symplectomorphisms $X\to X$, and to the
construction of quasi-modes for Toeplitz operators (both in progress).

\medskip
Our methods use heavily the machinery of Fourier integral operators
of Hermite type, developed by Louis Boutet de Monvel and Victor
Guillemin in \cite{BG}.  In fact, we associate to
closed Legendrian submanifolds of a strictly pseudoconvex domain,
$P$, distributions in the generalized Hardy space of $P$ (see \S 2).
The Szeg\"o projector is an Hermite FIO, and we show that our
Legendrian distributions posses a symbol calculus inherited from
that of Hermite distributions (symplectic spinors).

\medskip
The sections $u_k$ are defined as follows.
Let $P\subset L^*$ the unit circle bundle in the dual of $L$.
We denote by $\alpha$ the connection form on $P$; then the pair $(P,
\alpha)$ is a contact manifold and so it has a natural volume form,
\be
dp={\alpha\over 2\pi}\wedge (d\alpha)^n.
\ee
The disk bundle in $L^*$ is a strictly pseudoconvex domain;
we will consider the Hardy space of $P$, $\calS\subset L^2(P)$, and the
Szeg\"o projector
\be\label{1.2}
\Pi : L^2(P)\to \calS
\ee
given by orthogonal projection onto $\calS$.  The natural action of
$S^1$ on $P$ commutes with $\Pi$, and hence $\calS$ decomposes as
a Hilbert
space direct sum of isotypes.  Only positive frequencies arise in the
decomposition, and in fact the $k$-th summand is naturally identified
with $\calS_k$.  Therefore we identify
\be\label{1.3}
\calS\,=\,\bigoplus_{k=0}^\infty \calS_k\,.
\ee
Since we will use the calculus of Hermite Fourier integral operators,
we will actually need a metalinear structure on $P$.
This is a way of keeping track of the Maslov factors.

\medskip
Let $\Lambda\subset P$ be a compact Legendrian submanifold, and
$\nu$
a half-form on $\Lambda$.  It turns out that $\Pi$ extends to a class of
distributions including the delta function defined by $(\Lambda,
\nu)$.  We will suppress $\nu$ from the notation, and denote the
latter by $\dL$.

\begin{definition}
For each $k$, we denote by $u_k$ the $k$-th component
of $u:=\Pi(\dL)$ in the decomposition (\ref{1.3}).
\end{definition}

\noi{\bf Remarks:}
1.- Instead of a delta function along $\Lambda$ one can equally take
a conormal distribution to $\Lambda$, but to leading order asymptotics
the resulting states are not more general.
\par\smallskip\noi
2.- We regard the sequence $\{u_k\}$ as being associated with the
immersed Lagrangian $\Lambda_0:=\pi(\Lambda)$, where $\pi: P\to
X$ is
the projection.  Not all immersed Lagrangians in $X$ are of this
form; these are the {\em Bohr-Sommerfeld Lagrangians}.
\par\smallskip\noi
3.- In case the restriction of $\pi | : \Lambda\to \Lambda_0$ is a
covering
map with deck transformation group the group of $k_0$ roots of unity,
and the density $\nu$ is chosen invariant under it, then the Fourier
coefficients $u_k$ will be zero unless $k_0$ divides $k$.

\newcommand{\phik}{\varphi^{(k)}}
\medskip
The matrix element estimate, Theorem \ref{mainthm}, gives in
particular the asymptotics of the $L^2$ norms $\|u_k\|$.  The
resulting estimate can be
explained rather simply as follows.  For every $p\in P$, let
\be\label{1.4}
\phik_p\, :=\, \Pi_k(\delta_p)
\ee
be the orthogonal projection of the delta function at $p$ into
$\calS_k$.
(In case $X$ is a coadjoint orbit of a Lie group
these are the ``coherent states" of the Physicists.)
That is, if $\calK_k(q,p)$ is the Schwartz kernel of the orthogonal
projection $\Pi_k$,
\be\label{1.5}
\phik_p(q)\, :=\, \calK_k(q,p)\,,
\ee
and the reproducing property follows:
\be\label{1.6}
\forall f\in\calS_k\,\quad f(p)\,= \ip{f\,,\,\phik_p}\,.
\ee
Applying this to $f=\phik_p$ itself gives
\be\label{1.7}
\calK_k(p,p)\,=\, \| \phik_p\|^2\,,
\ee
and so
\be\label{1.8}
\mbox{dim }\calS_k\,=\,\int_P\,\calK_k(p,p)\,dp\,=\,
\int_P\,\|\phik_p\|^2\,dp\,.
\ee
By Riemann-Roch, we know this is a polynomial in $k$ of degree $n$
and
leading term  $(2\pi)^{-n} \mbox{Vol} (X)\, k^n$.
Thus we get that on average $\| \phik_p\|^2$ is of size $(2\pi)^{-n}
k^n$.
If we assume that $\|\phik_p\|^2$ is independent of $p$ (true for
example
if there is a transitive symmetry group present), then we actually get
\be\label{1.9}
\|\phik_p\|^2\, = (2\pi)^{-n} \, k^n\ + \mbox{ l. o. t.}
\ee

\medskip
On the other hand, by definition
\be\label{1.11}
u_k\, =\, \int_\Lambda\,\phik_p\, \nu_p\,,
\ee
and so the square of the norm is
\be\label{1.12}
\ip{u_k\,,\,u_k}\,=\,\int\int_{\Lambda\times\Lambda}\,
\ip{\phik_p\,,\,\phik_q}\, \nu_p\,\overline{\nu_q}\,.
\ee
{\em It turns out that this integral can be estimated for large $k$ by the
method of stationary phase.  For this it is crucial that $\Lambda$ be a
Legendrian submanifold.} The relevant critical points are on the
diagonal, $p=q\in\Lambda$, which is a non-degenerate manifold
of critical points.  Since the dimension of $\Lambda$ is $n$, we
see from (\ref{1.9}) that we should have
\be\label{1.13}
\ip{u_k\,,\,u_k}\,\sim\,\left({k\over \pi}\right)^{n/2}\,
\int_\Lambda\,|\nu|^2\,.
\ee
We will prove that this is indeed the case.

\medskip\noindent
{\sc Acknowledgments.}  We thank Svetlana Katok for advice and
encouragement regarding the application to Poincar\'e series.
Some of the calculations of \S4 were done in the Summer of
'93 jointly with J.J. Carroll, under the NSF sponsored REU program.

\newcommand{\inv}{^{-1}}
\newcommand{\bs}{\backslash}
\newcommand{\Ls}{\Lambda^{\sharp}}
\newcommand{\Lf}{\Lambda^{\flat}}
\newcommand{\hatch}{\mathbin{\#}}
\newcommand{\Zs}{\calZ^{\sharp}}
\newcommand{\Zf}{\calZ^{\flat}}
\renewcommand{\por}{{\cdot}}
\newfont{\eul}{eusm10 scaled\magstep1}
\newcommand{\eulS}{\mbox{\eul S}}
\mathchardef\perpchar="023F
\newcommand{\prp}{^\perpchar}
\newcommand{\Mp}{\mbox{Mp}}
\newcommand{\spin}{\mbox{\rm Spin}}
\newcommand{\hffm}{{\textstyle \bigwedge^{1/2}}}
\newcommand{\nhffm}{{\textstyle \bigwedge^{-1/2}}}
\newcommand{\diag}{\mathbin{\rlap{$\scriptstyle\times$}
        \raise.9ex\hbox{$\scriptscriptstyle\triangle$}}}
\newcommand{\heis}{\mbox{heis}}
\newcommand{\vol}{\hbox{\rm vol}}
\newcommand{\boxtimes}{\mathbin{\rlap{$\times$}
        \hskip.27ex\hbox{\vrule\vbox{\hrule\hbox{\phantom{e}}
        \hrule}\vrule}}}
\newcommand{\Ker}{\mathop{\mbox{\rm Ker}}}
\newcommand{\ycz}{Y\cap Z}

\section{Legendrian distributions and their symbols.}

In \cite{BG} Boutet de Monvel and Guillemin associate
spaces of distributions $I^m(\calM, \Sigma)$ on $\calM$ (called Hermite
distributions) to a conic
closed isotropic submanifold $\Sigma\subset T^*\calM\setminus\{ 0\}$.
In case $\Sigma$ is Lagrangian, these distributions
are precisely the classical Lagrangian distributions of H\"ormander
except that amplitudes of elements in $I^m(\calM, \Sigma)$ have asymptotic
expansions decreasing by half-integer powers of the fibre variables.
(There is also a discrepancy in the definition of order; we will follow
the conventions in \cite{BG}.)  Elements in $I^m(\calM, \Sigma)$ have
symbols, which are symplectic spinors on $\Sigma$.  Boutet de Monvel
and Guillemin prove a series of composition theorems regarding Hermite
distributions.  We will review this material as needed.

\subsection{The definition}

For all of \S 2 the setting is the following.
Let $P$ be a strictly pseudoconvex domain in a Stein manifold,
and let $\alpha$ be the pull-back to $P$ of $\Im \overline{\partial}\rho$,
where $\rho$ is a defining function for $P$.  Then $(P,\alpha)$ is a contact
manifold, and at every point $p$ the null space of $\alpha$ is the maximal
complex subspace of $T_pP$.  The null space has a symplectic structure,
namely
the one induced from $d\alpha$; therefore it has an associated Hermitian
structure.  Thus $P$ has a so-called pseudo-Hermitian structure.
Denote by $\calH\subset L^2(P)$ the Hardy space and
let $\Pi : L^2(P)\to\calH$ be the Szeg\"o projector.
Let $\Lambda\subset P$ be a closed Legendrian submanifold.

\begin{definition}
The space of Legendrian distributions of order $m$ associated with
$\Lambda$ is defined to be
\[
J^m(P,\Lambda)\, =\, \Pi\, (I^{m+n/2}(P, N^*\Lambda))\,.
\]
\end{definition}

Here $I^*(P, N^*\Lambda)$ denotes the spaces of conormal distributions
to $\Lambda$.  We must justify this definition; that is, we must show
that $\Pi$ extends to distributions conormal to $\Lambda$.

\medskip
Define a submanifold $\Zs\subset T^*P$ by
\be
\Zs := \{(p,r\alpha_p); p\in P, r>0\},
\ee
where $\alpha$ is the connection form on $P$.
$\Zs$ is in fact a symplectic submanifold of $T^*P$.

\begin{theorem}\label{szego}(\cite{BG}, Thm. 11.1)
Let
\be\label{2.0}
\calZ = \{\, (p, r\alpha_p; p, -r\alpha_p)\,;\,r>0\ \ {and}\ p\in P\,\}\,,
\ee
where $\alpha$ is the connection form on $P$.
Then $\calZ$ is an isotropic submanifold of $T^*(P\times P)$, and the
Schwartz kernel of $\Pi$ is an Hermite distribution in the space
$I^{1/2}(P\times P,\calZ)$.
\end{theorem}
We will use various composition
theorems of \cite{BG}.
For completeness, we make a definition that encapsules the
hypotheses of all of these theorems (\cite{BG}, conditions (7.4)).

\begin{definition}\label{compose}
Let $P$ and $Q$ be manifolds, and $\Gamma\subset T^*(Q\times
P)\setminus
\{ 0\}$ and $\Sigma\subset T^*P\setminus\{ 0\}$ be two closed
homogeneous submanifolds.
We will say that $\Gamma$ and $\Sigma$ are composible iff the following
hold:
\begin{enumerate}
\item $\Gamma$ should not contain vectors of the form $(q,\mu; p,0)$.
\item $\Gamma\circ\Sigma$ should not contain zero vectors.
\item If $\Gamma_0$ is the projection of $\Gamma$ onto $Q\times P$,
then
the projection $\Gamma_0\to Q$ is proper.
\item The fiber product
\be\label{2.01}
\begin{array}{ccc}
\calF &\to &\Gamma \\
\downarrow & & \downarrow\rho\\
\Sigma & \hookrightarrow & T^*P
\end{array}
\ee
where the right vertical arrow $\rho$
is the obvious projection, is {\em clean}.
\item The map $\tau : \calF\to T^*Q$ defined as the composition of the top
arrow in (\ref{2.01}) and the projection $\Gamma\to T^*Q$ is of constant
rank.
\end{enumerate}
\end{definition}
\noindent
We now recall the two conditions which define the cleanness of
a fiber product such as (\ref{2.01}).  The first requirement is that
$\calF$, which by definition is
\be\label{2.02}
\calF\, =\, \{\,(\gamma,\,\sigma)\in \Gamma\times\Sigma\, ;\,
\rho(\gamma)=\sigma\,\}\,,
\ee
is a submanifold of $\Gamma\times\Sigma$.
In addition, we require that for all $(\gamma, \sigma)\in\calF$,
\be\label{2.03}
T_{\gamma,\sigma}\calF\, =\, d\rho_\gamma^{-1}(T_\sigma\Sigma)\,.
\ee
The following integer plays an important role in the calculations:
\begin{definition}
The excess of the diagram (\ref{2.01}) is
\[
e\,=\,\dim (\calF) + \dim (T^*P) - \dim (\Gamma) - \dim (\Sigma)\,.
\]
\end{definition}
The geometrical meaning of the clean intersection is this: that
locally near every point in $\calF$ there is a submanifold
of $T^*M$ of codimension $e$, containing neighborhoods of the point in the
intersecting manifolds, which intersect transversely in the submanifold.

\bigskip
Now consider
\be\label{2.1}
\Ls := \Zs \cap N^*\Lambda
\,=\,\{\,(p,r\alpha_p)\,;\, r>0\ \mbox{and}\ p\in\Lambda\,\}\,,
\ee
where the second equality follows from $\Lambda$ being Legendrian.
$\Ls$ is a submanifold of the
conormal bundle of $\Lambda$ and hence is an isotropic submanifold
of $T^*P$.  It is a Lagrangian submanifold of $\Zs$.

\begin{proposition}\label{2.one}
The Szeg\"o projector extends by continuity to the space of
distributions on $P$ conormal to $\Lambda$. The extension maps
$I^{m}(P, N^*\Lambda)$ to the space $I^{m-n/2}(P, \Ls)$.
\end{proposition}
\begin{proof}
We apply Theorem 9.4 in \cite{BG}, which in the present case
says that if $\calZ$ and $N^*\Lambda$ are composible (in the sense of
Definition \ref{compose}), then an Hermite FIO
associated with $\calZ$ can be applied to a Lagrange distribution
associated with $N^*\Lambda$, and the result is an Hermite distribution
associated with $\Ls$.  Therefore all we need to do is to check
that $\calZ$ and $N^*\Lambda$ are composible, which is straightforward.
(The excess of the composition diagram turns out to be equal to $n$.)
\end{proof}

By Proposition (\ref{2.one}) not only the spaces $J$ are
well-defined, but in fact one has the inclusion
\be\label{2.2}
J^m(P,\Lambda)\subset I^m(P, \Ls)\, .
\ee
Although we won't need it here, we mention that these distributions
can also be described as ``marked Lagrangian distributions" in the
sense of Melrose, \cite{Me}, associated to the conormal bundle of
$\Lambda$ marked by the submanifold $\Ls$.

\subsection{The symbols of Legendrian distributions}
Our next task is to to show that one can
identify the symbol of an element in $J^m(P,\Lambda)$
with a half-form on $\Lambda$.
The symbol of a Lagrangian distribution is a half form on the Lagrangian
submanifold.   The symbol of an Hermite distribution is a more complicated
object, a symplectic spinor.   For the sake of completeness, we review
briefly the construction of symplectic spinors.

To any symplectic vector space $V$ there is naturally associated a Heisenberg
Lie algebra, denoted by $\heis(V)$, which as a vector space is just $V\oplus
\bbR$.   The Stone-von Neumann theorem gives us a unitary representation
$\rho$ of the associated Heisenberg group on a Hilbert space $H(V)\cong
L^2(\bbR^{\dim V/2})$.
If $V$ carries a metaplectic structure, then we can use the action of
$\mbox{Sp}(V)$ on $\heis(V)$ to construct a unitary representation of
$\Mp(V)$ on $H(V)$, the Segal-Shale-Weyl representation.
Denote by $H_\infty(V)$ the space of smooth vectors for this representation,
which is identified with the Schwartz functions $\eulS(\bbR^{\dim V/2})$.
Now let $Y\subset V$ be an isotropic subspace of dimension $k$.  Then
$Y\prp/Y$ inherits a
symplectic structure from $V$.  Moreover, the metaplectic structure on $V$
gives us the product of a metalinear structure on $Y$ and a metaplectic
structure on $Y\prp/Y$.  The space $\spin(Y)$ is defined by
\be
\spin(Y) := H_\infty(Y\prp/Y) \otimes \hffm Y.
\ee

Let $H_\infty(V)'$ denote the topological dual to
$H_\infty(V)$, which is identified with the space of tempered distributions
$\eulS'(\bbR^{\dim V/2})$.  The representation $d\rho$ of $\heis(V)$ on
$H(V)$ extends to a representation on $H_\infty(V)'$.  Identifying a
subspace $Y\subset V$ as a subalgebra of $\heis(V)$, we define
\be
\ker d\rho(Y) = \{f\in H_\infty(V)': d\rho(u)f=0\quad\forall u\in Y\}.
\ee
\begin{theorem}\label{kostant}
(Kostant)
For a Lagrangian subspace $Y\subset V$, the space
$\ker d\rho(Y)$ is one-dimensional and isomorphic to $\hffm Y$.
\end{theorem}

The bundle of symplectic spinors is defined as follows.   Let $M$ be a
metalinear manifold (a manifold possesses a
metalinear structure whenever the square of the first Stiefel-Whitney class
vanishes).   The choice of a metalinear structure on $M$ gives a canonical
metaplectic structure on $T^* M$.   Now let $\Sigma\subset T^* M$ be
an isotropic subspace.  Let $\Sigma_x$ denote the tangent space to $\Sigma$
at the point $x$.  The symplectic normal bundle, $\Sigma\prp/\Sigma$, is
the bundle whose fiber at $x$ is the space $\Sigma_x\prp/\Sigma_x$, where
$\Sigma_x\prp$ is the perpendicular space to $\Sigma_x$ in $T_x(T^*
M)$.    As above the metaplectic structure on $T^* M$ gives us the product
of a metalinear structure on $\Sigma$ with a metaplectic structure on
$\Sigma\prp/\Sigma$.  We can now define
$\spin(\Sigma)$ to be the vector bundle on $\Sigma$ whose fiber at a point
$x$ is given by
\be
\spin(\Sigma)_x = H_\infty(\Sigma_x\prp/\Sigma_x) \otimes \hffm
\Sigma_x.
\ee
Note that the fiber of $\spin(\Sigma)$ is an infinite dimensional vector
space.

A symplectic spinor on $\Sigma$ is a smooth section of $\spin(\Sigma)$.
There is an action of $\bbR^+$ on $\spin(\Sigma)$ coming from the action
of $\bbR^+$ on $T^* M\bs\{0\}$ given by $r\cdot(x,\xi)\mapsto
(x,r\xi)$.  Denote by $SS^m$ the space of symplectic spinors which are
homogeneous of degree $m$ under this action.
\begin{proposition}
(\cite{BG},  Prop. 7.4)
There is a canonical symbol map,
\be
I^m(M,\Sigma) \to SS^m(\Sigma),
\ee
whose kernel is $I^{m-1/2}(M,\Sigma)$.
\end{proposition}

In the present case, all of the relevant metalinear and metaplectic
structures derive from the choice of a metalinear structure on $P$.
In particular, $T^*P$ inherits a metaplectic structure, as do the horizontal
subspaces of $TP$.  These metaplectic structures on the horizontal subspaces
of $TP$ in
turn give rise to metalinear structures on Legendrian submanifolds of $P$.

\begin{lemma}
A pseudo-hermitian manifold $P$ always possesses a metalinear structure.
\end{lemma}
\begin{proof}
Let $w_1$ be the first Stiefel-Whitney class of $TP$,  an element
of $H^1(P,\bbZ/2)$.  The obstruction to the existence of a metalinear
structure is $w_1^2$.  Now $TP$ is orientable if and only if $w_1 = 0$.  Note
that this would take care of the case where $P$ is a circle bundle over $X$
as in \S1.
In general, we have
\be
TP = H\oplus E
\ee
(a Whitney direct sum of bundles), where $H$ is the horizontal distribution
(the kernel of $\alpha$), and $E$ is the trivial rank-one bundle spanned by
$\partial_\theta$.   By the axioms of Stiefel-Whitney classes, $w_1 =
w_1(H)$.   $H$ is a complex vector bundle, so in fact $w_1(H)=0$ since
a complex bundle is always orientable.
\end{proof}

\noi
A metalinear structure on $P$ is not necessarily unique.
The set of all metalinear structures on $P$ has the same cardinality as
$H^1(P, \bbZ/2)$.

\medskip
We will next describe the symbol of the Szeg\"o projector, $\Pi$.
Define $\Zs\subset T^* P$ as in \S2.1, and
$\Zf$ by
\be
\Zf  = \{(p,- r\alpha_p): p\in P, r>0\},
\ee
and note that the space $\calZ$ is the diagonal subspace
$\Zs_+\diag \Zf_-$.   Note that $\Zs$ and $\Zf$ are symplectic
submanifolds,  whereas $\calZ$ is an isotropic submanifold.    From Theorem
\ref{szego}, the Schwarz kernel of $\Pi$ is an Hermite distribution in
$I^{1/2}(P\times P,\calZ)$, so $\sigma(\Pi)$ is an element of
$SS^{1/2}(\calZ)$.

Since it is sufficient to describe $\sigma(\Pi)$ locally, we begin by
linearizing the problem.   Choose a point $(p,r\alpha_p)\in \Zs$ and let
$V=T_{(p,r\alpha_p)}(T^* P)$.  Define
\be
Z = T_{(p,r\alpha_p)}\Zs,
\ee
which is a symplectic subspace of $V$.  As vector spaces we can identify $V$
with $T_{(p,-r\alpha_p)}(T^* P)$ and $Z$ with
$T_{(p,-r\alpha_p)}\Zf$, but then $V$ and $Z$ carry the opposite symplectic
structures.  To avoid notational complications, we will use $V$ and $Z$ to
denote both of the respective vector spaces and point out the differing
symplectic structures where necessary.   We therefore write
\be
T_{(p,r\alpha_p,p,-r\alpha_p)}\calZ = Z\diag Z.
\ee
Note that $Z\diag Z$ is isomorphic to $Z$ but is an isotropic subspace of
$V\times V$.
At the point $(p,r\alpha_p, p, -r\alpha_p)$, the fiber of $\spin(\calZ)$ is
\be
\spin(Z\diag Z) = H_\infty((Z\diag Z)\prp/(Z\diag Z)) \otimes \hffm
(Z\diag Z).
\ee
\begin{lemma}\label{spinzz}
We have the canonical identification
\be
\spin(Z\diag Z) = H_\infty(Z\prp)\otimes H_\infty(Z\prp) \otimes
\hffm Z.
\ee
\end{lemma}
\begin{proof}
The spaces of half forms are identified through the isomorphism between $Z$
and $Z\diag Z$.
By definition $(Z\diag Z)\prp$ is the space of all $(v,w)\in V\times V$ such
that  $w-z\in Z\prp$.   Since $V=Z\oplus Z\prp$, it is clear that $(Z\diag
Z)\prp = (Z\diag Z) \oplus (Z\prp\times Z\prp)$.  Thus we have
\be
(Z\diag Z)\prp/(Z\diag Z) \cong Z\prp\times Z\prp,
\ee
which is a symplectic isomorphism.
\end{proof}

\medskip
The pseudo-Hermitian structure of $P$
enters into the description of $\sigma(\Pi)$ in
the following.
\begin{proposition}\label{kahler}
Associated to the pseudo-Hermitian structure of $P$ is a positive
definite Lagrangian subspace of $Z\prp\otimes \bbC$.
\end{proposition}
\begin{proof}
A K\"ahler structure on a vector space $V$ is equivalent to the combination
of a symplectic structure on $V$ and the choice of a positive definite
Lagrangian subspace of $V\otimes\bbC$ (the type $(1,0)$ subspace).  Thus we
need to show only that $Z\prp$ inherits a K\"ahler structure.
This follows from:

\medskip\noi
{\bf Claim:}
{\em Under the projection $T_{(p,r\alpha_p)}(T^* P)\to T_pP$,
the image of $Z\prp$ is the null space of $\alpha$ in $T_pP$.}

Indeed, let $s:P\times \bbR^+\to T^* P$ be the map $(p,r) \mapsto (p,
r\alpha_p)$, whose image is $\Zs$.  Then $Z$ is the image of the differential
map $ds$ at the point $(p,r)$ singled out above.  Explicitly,
\be
ds_{(p,r)}(v_r, v_p) = (v_p, v_r \alpha_p + rd\alpha_p(v_p)).
\ee
We quickly see that the perpendicular space to $Z$ is given by
\be
Z\prp = \bigl\{(w, -r(\nabla_p\alpha,w)):\>(\alpha_p, w) =0\bigr\},
\ee
from which the Claim follows.
\end{proof}

\begin{proposition}\label{4.2}
(\cite{BG},  Prop. 4.2)
If $W$ is a positive definite Lagrangian subspace of $Z\prp\otimes\bbC$,
then $\ker d\rho(W)\subset H_\infty(Z\prp)'$ is one-dimensional and
contained in $H_\infty(Z\prp)$.
\end{proposition}

Combining Propositions \ref{kahler} and \ref{4.2},
the pseudo-Hermitian structure on $P$ determines a one-dimensional
subspace of
$H_\infty(Z\prp)$.  To write the symbol of $\Pi$, we need to choose an
element $e$ of norm one in this space (the symbol is of course independent
of the choice).  For our purposes it is convenient to fix a particular
choice of $e$.  We can do this because $Z\prp$ is a symplectic vector space
with a K\"ahler structure.  Therefore there is a canonical realization of
the metaplectic representation $H(Z\prp)$ on Bargmann space.
We require that $e$ be real and positive under this representation.
\begin{theorem}\label{sszego}
(\cite{BG}, Thm. 11.2)
The symbol of the Szego projector is
\be
\sigma(\Pi) = e\otimes \overline{e}\otimes \sqrt{\vol_Z} \in
H_\infty(Z\prp)\otimes H_\infty(Z\prp)\otimes \hffm Z,
\ee
where $\vol_Z$ is the canonical volume form on $Z$ given by the
symplectic structure.
\end{theorem}

\bigskip
We now will describe the symbol of a Legendrian distribution, as a
symplectic spinor.
Let $\calK$ denote the Schwartz kernel of $\Pi$, and define the maps:
\be
\begin{array}{ccc}
P\times P&{\buildrel \Delta\over\longrightarrow}
&P\times P\times P\\ \pi\downarrow\phantom{\pi}&&\\
P&&\\
\end{array}
\ee
where $\Delta:(p_1,p_2) \mapsto (p_1,p_2,p_2)$ and $\pi:(p_1,p_2)\mapsto
p_1$.  Let $\zeta$ be a distribution on $P$ conormal to $\Lambda$.
Then $u = \Pi(\zeta)$ is given by
\be
u = \pi_*\Delta^* (\calK\boxtimes \zeta).
\ee
$\calK\boxtimes \zeta$ is an Hermite distribution on $P\times P\times
P$ whose isotropic relation is
$\calZ\times N^*\Lambda \subset T^*(P\times
P\times P)$.  The operator $\pi_*\Delta^*$ is an ordinary FIO with
associated
relation
\be
\calB = \{(p_1,\xi_1), (p_1,p_2,p_2, -\xi_1,\xi_2,-\xi_2)\} \subset T^*P
\times T^*(P\times P\times P).
\ee

Once again we proceed by linearizing the problem.   Fix a point
$(p,\xi)\in \Ls$, and define $V$ and $Z$ as in \S4.1.  Define
\be
B = T_{((p,\xi),(p,p,p,-\xi,\xi,-\xi))}\calB
\ee
We identify $B$ as a subspace of $V\times W$, where $W= V\times
V\times V$,
keeping track of the signs of the symplectic forms as needed.   We further
define
\be
Y = T_{(p,\xi)} (N^*\Lambda),
\ee
and
\be
A = (Z\diag Z)\times Y.
\ee
Our starting point for the calculation of $\sigma(u)$ is
$\sigma(\calK\boxtimes \zeta)\in\spin(A)$.  Because $Y$ is Lagrangian,
\be
A\prp/A \cong (Z\diag Z)\prp/(Z\diag Z) .
\ee
Using Lemma \ref{spinzz}, we can can thus canonically identify
\be
\spin(A) = H_\infty(Z\prp)\otimes H_\infty(Z\prp) \otimes \hffm
Z\otimes \hffm Y.
\ee

Denote the symbol of $\zeta$ by $\mu\in\hffm N^*\Lambda$.  This
corresponds to
an element of $\hffm Y$ which we also denote by $\mu$.
According to Theorem \ref{sszego}, we therefore have
\be
\sigma(\calK\boxtimes \zeta) = e\otimes \overline{e}\otimes
\sqrt{\vol_Z}\otimes \nu\,.
\ee
The symbol of $u$ will be an element of $\spin(\Ls)$.   The linearization of
$\Ls$ is $T_{(p,\xi)}\Ls = \ycz$.  As remarked in \S2.1, $\Ls$ is a
Lagrangian
submanifold of $\Zs$, and thus $\ycz$ is a Lagrangian subspace of $Z$.
\begin{lemma}
We have
\be
(\ycz)\prp/(\ycz) \cong Z\prp,
\ee
so that we can identify,
\be
\spin(\ycz) = H_\infty(Z\prp) \otimes \hffm (\ycz).
\ee
\end{lemma}
\begin{proof}
Since $\ycz$ is a Lagrangian subspace of $Z$ and $V=Z\oplus Z\prp$, we
have $(\ycz)\prp = Z\prp \oplus (\ycz)$.
\end{proof}
\begin{lemma}\label{sesqlemma}
For a fixed choice of $e\in H_\infty(Z\prp)$ there is a natural isomorphism
\be\label{new2.1}
\varphi_e: \hffm Y \to \hffm (\ycz).
\ee
\end{lemma}
\begin{proof}
We begin by noting that the direct sum decomposition $V = Z\oplus Z\prp$
induces the direct sum decomposition
\be
Y = (\ycz)\oplus (\ycz\prp).
\ee
Since $Y$ and $\ycz$ are Lagrangian subspaces of $V$ and $Z$ respectively,
$\ycz\prp$ must be a Lagrangian subspace of $Z\prp$.

We thus have
\be\label{hffmy}
\hffm Y \cong \hffm (\ycz)\otimes \hffm(\ycz\prp).
\ee
Theorem \ref{kostant} gives the identification
\be
\hffm(\ycz\prp) \cong \Ker d\rho(\ycz\prp) \subset H_\infty(Z\prp)'.
\ee
Thus the Hilbert space inner product extends to a pairing
\be
H_\infty(Z\prp)\otimes \hffm(\ycz\prp) \to \bbC,
\ee
which, combined with (\ref{hffmy}), yields the map $\varphi_e$.

To see that $\varphi_e$ is an isomorphism, note that the unitary
group $U(Z\prp)\subset \mbox{Sp}(Z\prp)$ acts transitively on the set of
Lagrangian subspaces of $Z\prp$, while preserving the K\"ahler structure.
Thus we can choose an identification $Z\prp \cong \bbR^{2n}$ such that
$e\in\eulS(\bbR^n)$ is a Gaussian centered at the origin, and $\Ker
d\rho(\ycz\prp)$ consists of constant multiples of the delta function at the
origin.  Therefore (\ref{new2.1}) is an isomorphism.
\end{proof}

\begin{proposition}\label{legsym}
As a symplectic spinor, the symbol of $u = \Pi (\zeta)$ is
\be
\sigma(u) = e \otimes \varphi_e(\mu) \in H_\infty(Z\prp) \otimes
\hffm(\ycz)
\ee
(which is independent of the choice of $e$).
\end{proposition}
\begin{proof}
Consider $B\subset V\times W$ as a canonical relation from $W$ to $V$:
\be
\begin{array}{lcr}
&B\\
\raise.6ex\hbox{$\alpha$}\swarrow&&
\searrow\raise.6ex\hbox{$\beta$}\\
W&&V\\
\end{array}
\ee
where $\alpha$ and $\beta$ are the obvious projections.  The result of the
composition is $B\circ A := \beta(\alpha\inv(A)) = Y\cap Z$.  Proposition
6.5 of \cite{BG} gives, in the present case, the symbol map
\be\label{prop6.5}
\spin(A) \otimes \hffm B \to \spin(B\circ A).
\ee

The construction of the map (\ref{prop6.5}) has two essential components.
The first is
an exact sequence
\be\label{tseq}
0 \to  \Ker \rho \to B\oplus A \;{\buildrel \tau\over\to}\; U_1\prp \to
0,
\ee
where $\tau:B\oplus A \to W$ is defined by $((a,b),c) \mapsto b-c$, and
$U_1 = \alpha(B)\prp \cap A\prp \subset W$.   A simple computation
reveals $U_1\cong  \ycz\prp$.  In our case, $\Ker \rho \cong B\circ
A = \ycz$, so that this exact sequence gives an isomorphism
\be\label{hfiso}
\hffm B \otimes \hffm Z \otimes \hffm Y
\cong \hffm (U_1\prp) \otimes \hffm (\ycz)
\ee
Let $U$ be the image of $U_1$ in the quotient $A\prp/A$ (note that $U\cong
U_1$).   Under the identification of $A\prp/A$ with $Z\prp\times Z\prp$,
$U$ is just given by to $\{0\}\times (\ycz\prp)$.

The other component of the map (\ref{prop6.5}) is the isomorphism
(\cite{BG}, 4.15)
\be\label{kdpu}
\Ker d\rho(U) \cong \hffm U \otimes H_\infty(U\prp/U)'.
\ee
It is clear from the remarks above that $U\prp/U = Z\prp\times\{0\}$.
Since $\Ker d\rho(U)\subset H_\infty(A\prp/A)'$, taking the dual of the
isomoprhism (\ref{kdpu}) gives a map
\be\label{dlmp}
H_\infty(Z\prp)\otimes H_\infty(Z\prp) \otimes\hffm (\ycz\prp) \to
H_\infty(Z\prp).
\ee
Note that this is not an isomorphism.

These components fit together as follows.  We begin with
\be
\spin(A) \otimes \hffm B = H_\infty(Z\prp) \otimes H_\infty(Z\prp)
\otimes \hffm A \otimes \hffm B.
\ee
 The isomorphism (\ref{hfiso}) takes us to
\be\label{ustage}
H_\infty(Z\prp) \otimes H_\infty(Z\prp) \otimes \hffm (U_1\prp)
\otimes \hffm (\ycz).
\ee
Now because of the natural isomorphism $U_1\prp \cong (W/U_1)^*$, we
have
\be
\hffm U_1\prp \cong  \nhffm W \otimes\hffm U.
\ee
$W$ possesses a canonical half-form, which gives us a map $\nhffm W\to
\bbC$, so that we can naturally identify
\be\label{uyp}
\hffm U_1\prp \cong \hffm (\ycz\prp).
\ee
Thus (\ref{ustage}) is isomorphic to
\be\label{lastiso}
H_\infty(Z\prp) \otimes H_\infty(Z\prp) \otimes \hffm(\ycz\prp)
\otimes \hffm (\ycz).
\ee
The map (\ref{dlmp}), the only stage which is not an isomorphism,
completes construction of the map (\ref{prop6.5}).

We now simply trace what happens to the combination of
$\sigma(\calK\boxtimes
\zeta) \in \spin(A)$ and $\sigma(\pi_*\Delta^*)\in \hffm B$ under this
map.  First of all, we note that $\pi_*\Delta^*$ is a naturally defined
operator, and $\sigma(\pi_*\Delta^*)$ is just the
canonical element of $\hffm B$ determined by the symplectic structure on $B
\cong V\times V$.

Consider the point (\ref{ustage}) in the construction of the map.    In the
present case,
\be\label{uprp}
U_1\prp \cong V\times V\times [Z\oplus (\ycz\prp)] \subset W,
\ee
so that the isomorphism
\be
\hffm B \otimes \hffm Z\otimes \hffm Y \cong \hffm (U_1\prp) \otimes
\hffm(\ycz)
\ee
simply arises from the decomposition $Y = (\ycz)\oplus (\ycz\prp)$.
In view of (\ref{uprp}), the isomorphism (\ref{uyp}) consists simply of
dividing out by
the canonical half-forms on $B$ and $Z$.  Since the symbol of
$\pi_*\Delta^*$ and the half-form part of the symbol of $\Pi$ are in fact
just the canonical half-forms, these cancel out.
At the stage (\ref{lastiso}) we end up with
\be
\sigma(\calK\boxtimes \zeta) \otimes  \sigma(\pi_*\Delta^*) = e \otimes
\ol{e} \otimes \mu,
\ee
where $\mu$ is thought of as an element of $\hffm(\ycz\prp)
\otimes \hffm (\ycz)$.
The final stage is to apply the map (\ref{dlmp}), which takes $e\otimes
\ol{e} \otimes \mu$ to $e\otimes \varphi_e(\mu)$.
\end{proof}

Observe that $\varphi_e(\mu)\in\hffm (T_{(p,r\alpha_p)}\Ls)$.
Letting $p$ and $r$ vary,
$\varphi_e(\mu)$ defines a half-form on $\Ls$ which is homogeneous.
In view of the previous results, this half-form is the non-trivial part
of the symbol of $u$, as an Hermite distribution.
Since $\varphi_e(\mu)$ is homogeneous, it is determined by the
restriction to the image of the section
\be\label{new2.2}
\begin{array}{rcc}
s_\alpha : \Lambda& \to & \Ls\\
p & \mapsto & \alpha_p
\end{array}
\ee
Upon division by the natural radial half-form, this restriction
becomes a half-form on $\Lambda$.  We will refer to this half-form
as the pull-back of $\varphi_e(\mu)$ to $\Lambda$ via $s_\alpha$
and denote it by $s_\alpha^* \varphi_e(\mu)$.

\begin{definition}
For a Legendrian distribution $u$, we will identify the symbol of $u$ with
the half-form on $\Lambda$, $s_\alpha^*\varphi_e(\mu)$.  Precisely, we call
the (well-defined) map
\[
\begin{array}{rcc}
\sigma^{(m)} : J^m(P,\Lambda) &\to & \hffm\Lambda\\
u=\Pi(\zeta) & \mapsto & s_\alpha^*\varphi_e(\mu)
\end{array}
\]
the symbol map of order $m$.
\end{definition}

\subsection{Exactness of the symbol sequence.}

Our goal here is to prove the following:

\begin{theorem}
The following is an exact sequence:
\be\label{n2.1}
0\to J^{m-1/2}(P,\Lambda)\to J^m(P,\Lambda)\to \hffm\Lambda\to 0\,.
\ee
Moreover this sequence has a natural splitting, namely
\be\label{n2.2}
\begin{array}{rcl}
\hffm\Lambda  & \to & J^m(P,\Lambda)\\
\nu & \mapsto & T^{m-{1\over 2}}\Pi (\delta_\nu)
\end{array}
\ee
where $T=\Pi \partial_T\Pi$ and $\partial_T$ is the contact vector field
(defined by the conditions $\iota_{\partial_T}\alpha = 1$ and
$\iota_{\partial_T}d\alpha=0$).
\end{theorem}

\noindent
{\bf Remark:}  The operator $T$ is non-negative.  This follows
from the fact that the symbol of $\partial_T$ restricted to
$\calZ$ is positive.  Then, as in Proposition 2.14 of \cite{BG}, there
exists a {\em non-negative, elliptic} self-ajoint pseudodifferential
operator, $A$, on $P$ such that
\be\label{n2.3}
[A\,,\Pi]\,=\,0\quad\mbox{and}\quad \Pi A\Pi\,=\, \Pi \partial_T\Pi\,
=\, T\,.
\ee
Therefore the powers $T^s= \Pi A^s\Pi$ ($A^s$ defined to be zero in the
kernel
of $A$) are Toeplitz operators of order $s$, $\forall s\in\bbR$.

\medskip
We first check that (\ref{n2.2}) is a right inverse of the symbol
map.  We need the following result on
the behavior of the spaces $J$ under Toeplitz operators:

\begin{lemma}
If $u\in J^m(P,\Lambda)$ and $S=\Pi B\Pi$ is a Toeplitz operator of order
$p$, then $S(u)\in J^{m+p}(P,\Lambda)$ and its symbol is
$s_\alpha^*(\sigma_B)\sigma(u)$.  In particular,
$\forall s\in\bbR$, $T^s$ maps $J^m(P,\Lambda)$ into
$J^{m+s}(P,\Lambda)$, and
this map is the identity at the symbol level.
\end{lemma}
\begin{proof}
By \cite{BG}, without loss of generality
$[B,\Pi]=0$.  Therefore $S(u)=\Pi B(u)$, and since $B(u)$ is
another conormal distribution the proof is complete.
\end{proof}

\begin{corollary}\label{n2split}
Indeed (\ref{n2.2}) is a right inverse of the symbol map.
\end{corollary}

\medskip
Next we prove that the kernel of the symbol map is precisely
$J^{m-1/2}(P,\Lambda)$.  The non-trivial part is to show that if
$u=\Pi(\zeta) \in J^m(P,\Lambda)$ has zero symbol of order $m$,
then it is the projection of a conormal distribution of order
ord$(\zeta)-1/2$.  This is a consequence of the following:

\begin{theorem}
\[
\bigl\{\,u\in I^{m}(P,\Ls)\;;\;\Pi(u) = u\,\bigr\}\>=\>
J^m(P,\Lambda)\quad \mbox{modulo smooth functions.}
\]
\end{theorem}
\begin{proof}
Suppose that $u\in I^{m}(P,\Ls)$ is invariant under $\Pi$.  Then the
symbol $\sigma_u$ of $u$ (as an Hermite distribution)
is a symplectic spinor which equals its own composition with the
symbol of $\Pi$.  From the discussion of the symbol of $\Pi$
one can see that this implies that $\sigma_u$ is of the form
\be\label{n2.4}
\sigma_u\, =\, e\otimes\mu\,,
\ee
where $\mu\in\hffm\Ls$.  By Corollary \ref{n2split} it is possible to
construct a
conormal distribution $\zeta_1\in I^{m+n/2}(P, N^*\Lambda)$ such that
$\Pi(\zeta_1)$ and $u$ are Hermite distributions with the same
symbol.  Therefore, by the general symbol calculus of \cite{GU},
\be
u_1\, :=\, u-\Pi(\zeta_1) \, \in \, I^{m-1/2}(P, \Ls)\,.
\ee
Observe furthermore that $\Pi(u_1) = u_1$; therefore we can repeat the
same argument with $u_1$.  Continuing by induction, we see that there
is a sequence of conormal distributions, $\{\zeta_j\}$ whose orders
are monotonically decreasing such that $\forall k\in\bbZ^{+}$
\be
u-\Pi\bigl(\sum_{j=1}^k\,\zeta_j\bigr)\,\in\,I^{m-k/2}(P,\Ls)\,.
\ee
Now let $\zeta$ be a conormal distribution such that
$\zeta\sim\sum_{j=1}^\infty\zeta_j$.  Then $u-\Pi(\zeta)$ is a smooth
function.  The converse inclusion is trivial.
\end{proof}

\section{Matrix Element Estimates}
\renewcommand{\Ls}{\Lambda^{\sharp}}
\renewcommand{\Lf}{\Lambda^{\flat}}
\renewcommand{\hatch}{\mathbin{\#}}
\renewcommand{\calZ}{{\cal Z}}
\renewcommand{\Zs}{\calZ^{\sharp}}
\renewcommand{\Zf}{\calZ^{\flat}}
\renewcommand{\por}{{\cdot}}
\newcommand{\del}{\partial}
\newcommand{\yocz}{{Y_1\cap Z}}
\newcommand{\ytcz}{{Y_2\cap Z}}
\newcommand{\tso}{{T_1S^1}}
\newcommand{\cso}{{T^*_1S^1}}
\newcommand{\mzs}{\backslash\{0\}}

For this section we return to the case described in \S1,
where $P$ is a unit circle bundle over a compact Kahler manifold $X$.
In \S2, we saw that at the symbolic level all Legendrian distributions look
like delta functions or their derivatives.  In view of this fact, we will
restrict ourselves to the delta function case for the sake of simplicity.
Let $\Lambda_1$ and $\Lambda_2$ be compact Legendrian submanifolds of
$P$, and define Legendrian distributions $u = \Pi(\delta_{\Lambda_1})\in
J^{1/2}(P,\Lambda_1)$ with symbol $\nu_1 \in\hffm \Lambda_1$ and $v =
\Pi(\delta_{\Lambda_2}) \in J^{1/2}(P,\Lambda_2)$ with symbol
$\nu_2\in\hffm \Lambda_2$.
Let $A$ be a zeroth order classical pseudodifferential operator
on $P$ and let $T_A = \Pi A\Pi$ be the corresponding Toeplitz operator.
In this section we estimate the matrix elements $\ip{T_A u_k\,,\,v_k} =
\ip{Au_k, v_k}$.

\subsection{The main statements}

Let $F: P\times S^1\to P$ be the action map, and define
\be\label{2.0a}
\Theta_2\,:=\,F^{-1}(\Lambda_2)\,=\,\{\,(p,\omega)\,;\,
p\por\omega\in\Lambda_2\,\}\,,
\ee
where $\omega \in S^1$ and the action is denoted by a dot.
$\Theta_2$ is a submanifold since $F$ is a submersion; in fact
$\Theta_2\cong \Lambda_2\times S^1$ by the map $(p,\omega)\mapsto
(p\por\omega,\omega)$.

\bigskip\noi
{\bf Assumption:} {\em We will assume that
$\Lambda_1\times S^1$ and $\Theta_2$ intersect cleanly,
meaning:
\begin{enumerate}
\item The intersection
\be\label{2.0a1}
\calP\, :=\, (\Lambda_1\times S^1)\cap\Theta_2\, =\, \{\, (p,\omega)\,
;\, p\in\Lambda_1\  \mbox{and}\ p\por\omega\in\Lambda_2\,\}
\ee
is a submanifold of $P\times S^1$.
Different connected components of $\calP$ are allowed to
have different dimensions.
\item At every $(p,\omega)\in\calP$,
\[
T_{(p,\omega)}\calP\,=\,T_{(p,\omega)}\Theta_2\cap
T_{(p,\omega)} (\Lambda_1\times S^1)\,.
\]
\end{enumerate}}

Equivalently, we may assume that the image of the map $\Phi : \calP\to
S^1$ induced
by the natural projection is finite, and that for $\omega\in \Phi(\calP)$, the
intersection $(\Lambda_1\cdot\omega) \cap \Lambda_2$ is clean.   That
this is equivalent to the above assumption follows from the fact that
$\Lambda_1$ and $\Lambda_2$ are Legendrian.

We label the points in the image of $\Phi$ by
\be\label{2.0b}
\Phi(\calP) = \{\,\omega_1,\,\ldots,\,\omega_N\,\}.
\ee
For each $l \in\{1,\ldots,N\}$ let
$d_l$ be the dimension of the fiber $\Phi^{-1}(\omega_l)$, i.e.,
\be
d_l = \dim (\Lambda_1\cdot \omega_l) \cap \Lambda_2.
\ee

\medskip
\begin{lemma}\label{legint}
If $\Lambda_1$ and $\Lambda_2$ are two cleanly intersecting
Legendrian submanifolds
of $P$ and $\mu_1$ and $\mu_2$ are half-forms on the
respective submanifolds, then the intersection $\Lambda_1\cap
\Lambda_2$ inherits a
top degree form $\mu_1 \hatch \mu_2$.
\end{lemma}
\begin{proof}
Let $Z$ by a symplectic vector space, with Lagrangian subspaces $L_1$ and
$L_2$.  The exact sequence,
\be
0 \to L_1\cap L_2 \to L_1\oplus L_2  \to L_1 +L_2 \to 0,
\ee
where the third arrow takes $(v,w)\mapsto v-w$, leads to an isomorphism
\be
\hffm L_1 \otimes \hffm L_2 \cong \hffm (L_1\cap L_2) \otimes \hffm
(L_1+L_2).
\ee
Now, since $(L_1\cap L_2)\prp = L_1+L_2$, we have
\be
L_1 + L_2 \cong  [Z/(L_1\cap L_2)]^*.
\ee
This in turn allows us to identify
\be
\hffm (L_1+L_2) \cong \hffm (L_1\cap L_2),
\ee
by using the canonical half-form on $Z$ to map $\nhffm Z \to \bbC$.
Finally, there is a natural map
\be
\hffm (L_1\cap L_2) \otimes \hffm (L_1\cap L_2)  \to \wedge^d (L_1\cap
L_2),
\ee
where $d = \dim (L_1\cap L_2)$.

This construction thus associates to a pair of half-forms on the Lagrangian
subspaces a top degree form on their intersection.   Clearly the procedure
generalizes to the case of two cleanly intersecting Lagrangian submanifolds
of a symplectic manifold.

Define $\Ls_1, \Ls_2$, and $\Zs$ as in \S2.1.  $\Zs$ is a symplectic
manifold, and $\Ls_1$ and $\Ls_2$ are Lagrangian submanifolds.  To a
half-form $\nu_j$ on $\Lambda_j$ we naturally associate a half-form on
$\Ls_j$ by
\be
\nu_j \mapsto \nu_j \otimes \sqrt{dr}.
\ee
The result proven above gives us a top degree form $(\nu_1\otimes
\sqrt{dr})\hatch (\nu_2\otimes \sqrt{dr})$ on $\Ls_1\cap
\Ls_2$.  We define $\nu_1\hatch\nu_2$  as this form divided by $\alpha$.
\end{proof}

\medskip
Proceeding by analogy with the Fourier integral operator calculus,
one might guess that the leading coefficient in the
estimates for the matrix elements would involve only
universal constants and the
natural pairing described in Lemma \ref{legint}.  In fact, this coefficient
involves an additional term, which we now describe.

Consider the tangent space $T_xP$ at a point $x\in P$.  In \S2 we noted that
the null space of $\alpha$ in $T_xP$ (the horizontal space) is a symplectic
vector space.  Thus we
can define an action of the symplectic group $Sp(n)$ on $T_xP$ by its action
on the null space of $\alpha$, and acting trivially on vertical vectors.
In a symplectic vector space the
unitary group, $U(n)$, regarded as a subgroup of $Sp(n)$, acts
transitively on the set of Lagrangian subspaces.  In our case $U(n)$
acts transitively on the set of tangent spaces (at $x$) to oriented
Legendrian submanifolds, with isotropy subgroup $SO(n)$.
Thus, given two oriented Legendrian subspaces
$\Lambda_1$ and $\Lambda_2$ of $P$, we have a well-defined function on
$\Lambda_1\cap\Lambda_2$ which is the determinant of the unitary
matrix mapping $T_x\Lambda_1$ to $T_x\Lambda_2$.  We denote this
function by $\det\{\Lambda_1,\Lambda_2\}$.
Alternatively, if positive
orthonormal bases $\{e_j\}$ and $\{f_j\}$ are chosen for
the respective tangent spaces at $x$, we may define
\be
\det\{\Lambda_1,\Lambda_2\}(x) = \det \{h(e_j, f_j)\},
\ee
where $h$ is the hermitian form (on the null space of $\alpha$) at $x$.

In what follows, we will need to make sense of the square root of this
function.  This is precisely the role of the metalinear and metaplectic
structures.  As we remarked in \S2, the metalinear structure on $P$ gives
rise to a metaplectic structure on the horizontal subspace of $TP$ and
to metalinear structures on the Legendrian submanifolds.  To a unitary
transformation as described above, we can associate a well-defined element
in the double
cover of $U(n)$ by taking the unique element of $Mp(n)$ which lies over the
given transformation and which is a metalinear map from the tangent
space of one Legendrian to the other.  This association allows us to
define the square root of $\det\{\Lambda_1, \Lambda_2\}$:
the function $\sqrt{\det}$ is well-defined on the double cover of $U(n)$.

\medskip
With these assumptions and notation, we are now prepared to state our main
result.
\begin{theorem}\label{mainthm}
As $k\to \infty$ there is an asymptotic expansion
\be\label{2.00}
\ip{T_Au_k\,,\,v_k}\,\sim\, \sum_{l=1}^N
\omega_l^k\sum_{j=0}^\infty\,
c_{j,l}\,k^{(d_l-j)/2}\,,
\ee
with
\be\label{2.0c}
c_{0,l}\,=\,\,2^{(n-d_l)/2} \pi^{-d/2}\,\int_{(\Lambda_1\cdot\omega_l)
\cap \Lambda_2} \det\{\Lambda_1\cdot\omega,
\Lambda_2\}^{-1/2}\;  a\nu_1 \hatch \ol{\nu_2},
\ee
where $a$ is the pull-back to $\Lambda_1$ of the symbol of $A$ by
the connection one-form, $\alpha$.  Furthermore, if $\pi(\Lambda_1)$ and
$\pi(\Lambda_2)$ do not intersect, then $\ip{T_Au_k\,,\,v_k}$ decreases
rapidly in $k$.
\end{theorem}

\noi
Consider now the case of a single $u \in J^m(P,\Lambda)$, with symbol
$\nu$.  Observe that $\Lambda\times \{1\}\subset \calP$ and is always a
component of maximal dimension, $n$.  Furthermore,  if
$\pi\,:\Lambda\to\pi(\Lambda)$ is a covering map with
covering group the group of $k_0$-th roots of unit, then
$\Lambda\times \{ e^{2\pi i j/k_0}\}$, $j=0,\ldots, k_0-1$, are
$n$-dimensional components as well.  This leads us to the following
corollary.

\begin{corollary}\label{maincor}
Let $\pi(\Lambda)$ be a covering map with covering group the group of
$k_0$-th roots of unity, and suppose the half-form $\nu$ on $\Lambda$ is
invariant under the action of the covering group.  Then $\ip{u_k,u_k}$ has
the asymptotic behavior:
\be
\ip{u_k,u_k} \sim  k_0 \left({k\over\pi}\right)^{n/2}\,
\,\int_{\Lambda} |\nu|^2
\ee
if $k_0$ divides $k$ ($u_k = 0$ otherwise).
In particular, for $k$ a sufficiently large multiple of $k_0$,
$\ip{u_k,u_k}$ is non-zero.
\end{corollary}

The remainder of \S3 is devoted to the proof of the asymptotic
expansion of Theorem \ref{mainthm}.
We begin by dealing with the case where the immersed Lagrangians
corresponding to the Legendrian submanifolds do not
intersect.  The strategy is to study the singularities of the
periodic distribution
\be\label{2.000}
\Upsilon(\theta)\,:=\,\sum_{k=0}^\infty\,\ip{Au_k\,,\,v_k}
\,e^{ik\theta}\,,
\ee
as in \cite{GU}.  Knowledge of the singularities of $\Upsilon$
translates into the asymptotic expansion of its Fourier coefficients.
In particular, we will show that the wave-front set of $\Upsilon$ is empty
when $\calP = \emptyset$.  In this case, $\Upsilon$ is smooth and the
matrix elements must decrease rapidly in $k$.

\medskip
To construct $\Upsilon$ we proceed as follows.   Choose
$\zeta\in I^{(n+1)/2}(P,N^*\Lambda_2)$, such that $v = \Pi(\zeta)$.  We
will demonstrate that
\be\label{2.001}
\calV := F^*(\ol{\zeta})
\ee
(well-defined because $F$ is a submersion), is a Lagrangian distribution on
$P\times S^1$.  Here $\overline{\zeta}$ is
the complex conjugate of the distribution $u$ as defined by the
identity
\[
(\overline{\zeta},\varphi)\,=\,
\overline{ (\zeta, \overline{\varphi})}\,.
\]
Furthermore,
\be\label{2.3}
\calV(p, e^{i\theta})\,
=\,\overline{\zeta(p\cdot e^{i\theta})}.
\ee
We can regard $\calV$ as the Schwartz
kernel of an operator, $V$, from $P$ to the circle.  Since $\Pi$ is an
orthogonal projection which has already been applied to obtain $u$, it is clear
that
\be\label{2.002}
\Upsilon\,=\,V\circ A (u),
\ee
independently of the choice of $\zeta$.

We begin by describing the
canonical relation of the standard FIO $F^*$.   If $\eta\in T^*_pP$, denote
by $\eta^{\circ}$ the horizontal part of $\eta$.  Thus we decompose
\[
\eta\,=\,\eta(\partial_\theta)\alpha_p + \eta^{\circ}\,.
\]
For every $\omega\in S^1$, we denote by $R_\omega :P\to P$ the
map induced by the action of $\omega$ on the right.  We define an operator
$\tilde R_\omega: T^*_{p\cdot\omega}P\to T^*_pP$ by
\be
\tilde R_\omega :\eta \mapsto \eta(\partial_\theta)\alpha_p +
R^*_\omega \eta^{\circ}
\ee
The canonical relation of the Schwartz kernel of $F^*$ is
\be\label{2.3.1}
\calC\,=\,\{\, ( p,\omega; \tilde R_\omega(\eta),
\eta(\partial_\theta))\,,\,
(p\por \omega;\eta)\,\}\, \subset T^*(P\times S^1)
\times T^*P.
\ee
The following result is well-known.
\begin{proposition}
The pull-back operator, $F^*$, extends to a map
\be\label{2.2.2}
F^*\, : I^{m}(P,N^*\Lambda_2)\to I^{m}(P\times S^1, \Gamma)\,,
\ee
where
\be\label{2.2.3}
\Gamma\,=\, \{\,(p, \omega\, ;\tilde R_\omega(\eta),
\eta(\partial_\theta))\,; (p\por\omega,\eta)\in N^*\Lambda_2\,\}\,
\subset
T^*(P\times S^1).
\ee
In particular, $\calV \in I^{(n+1)/2}(P\times S^1,\Gamma)$.
\end{proposition}

\bigskip
We can now consider $\calV$ as the Schwartz kernel of an operator,
$V$, from $P$ to the circle.  Specifically,
\be\label{2.5}
V(f)(\omega)\,=\,\int_P\,f(p)\,\overline{\zeta(p\cdot\omega)}\,dp\,.
\ee
We wish to apply $V$ to
$T_A(u)$, where $T_A$ is a Toeplitz operator.

\begin{proposition}\label{Upsmooth}
$\Upsilon = V (T_A (u))$, and
$\mbox{WF }(\Upsilon)\,=\,\bigcup_{j=1}^N \{ \omega_j\}\times\bbR^{+}$.
In particular,
if $\calP = \emptyset$ then $\Upsilon$ is smooth.
\end{proposition}
\begin{proof}
By Proposition 2.13 of \cite{BG}, we can assume that $[A,\Pi]=0$,
and so $T_A:I^m(P, \Ls_1) \to I^m(P, \Ls_1)$.  Therefore we need
only to apply e.g. Theorem 8.2.13 in \cite{Ho}.  We omit the details.
\end{proof}

\begin{corollary}
If $\pi(\Lambda_1)\cap \pi(\Lambda_2) = \emptyset$ then the matrix
elements $\ip{T_Au_k, v_k}$ decrease rapidly in $k$.
\end{corollary}

\medskip\noi
{\bf Remark.}
If $\calP$ is non-empty one can ask whether the above construction
shows that $\Upsilon$ is a Lagrangian distribution, i.e. whether
the composition fiber product diagram
\be\label{2.10}
\begin{array}{rcl}
\calG & \to & \Gamma\\
\downarrow & &\downarrow\\
\Ls_1 & \to & T^*P
\end{array}
\ee
(where
\be\label{2.11}
\calG\,=\,\{\, [(q,\omega,r\alpha_q, r),(p,r)]\,;
\,p=q\por\omega\,,\,p\in \calP\,\}
\ee
and the arrows are the obvious ones) is clean.  This is not so
because condition (\ref{2.03}) is not satisfied.  To proceed, we will
estimate the Fourier coefficients of $\Upsilon$ directly using the
stationary phase lemma.  This is the content of \S3.2 and \S3.3.

\subsection{Oscillatory integrals}

The strategy for computing the asymptotic estimates of \S3.1 is to write the
Hermite distributions as oscillatory integrals and approximate the matrix
elements by stationary phase.  By definition, the space $I^m(P, \Ls)$
consists of distributions that are locally expressible as oscillatory
integrals of a certain type.   For the remainder of this section, we will
be working with open neighborhoods in $P$ and $\Lambda$.   References to
$P$ and $\Lambda$ below are to be interpreted as statements concerning
local neighborhoods in these spaces (else the notation becomes excessively
complex).

In order to write the oscillatory integrals, we must first find a phase
function parametrizing $\Ls$, in a sense to be described below.  We need
only consider
the case of a non-degenerate phase function.  The set up is
as follows.  Let $B$ be an open conic subset of $(\bbR\times
\bbR^n)\mzs$. We give $\bbR\times \bbR^n$ the coordinates $(\tau,
\eta)$.
\begin{definition}
A non-degenerate phase function is a function $\phi\in C^\infty(P\times B,
\bbR)$ which satisfies:
\begin{enumerate}
\item $\phi$ is homogeneous in $(\tau,\eta)$.
\item $d\phi$ is nowhere zero.
\item The critical set of $\phi$,
\be
C_\phi = \{(x,\tau,\eta);  (d_\tau\phi)_{(x,\tau,\eta)} =
(d_\eta\phi)_{(x,\tau,\eta)} = 0  \},
\ee
intersects the the space $\eta_1 = \ldots = \eta_n = 0$ transversally.
\item The map $(x,\tau,\eta) \mapsto ({\del\phi\over \del \tau},
{\del\phi\over \del \eta_1}
\ldots,  {\del\phi\over \del \theta_{n+1}})$ has rank $n+1$ at every
point of $C_\phi$, i.e. $\phi$ is non-degenerate.
\end{enumerate}
\end{definition}

Define the map $F:C_\phi \to T^*P\mzs$ by $(x, \tau, \eta) \mapsto (x,
(d_x\phi)_{(x,\tau,\eta)})$.  We quote the following result.
\begin{proposition}
The image under $F$ of the subspace of $C_\phi$ given by
$\eta_1=\ldots=\eta_n = 0$ is a homogeneous isotropic submanifold of
$T^*P\mzs$ of dimension $n+1$.   \end{proposition}

\begin{definition}
A phase function $\phi$ is said to parametrize $\Ls$ provided
$\Ls$ is the image under $F$ of $C_\phi\cap \{\eta_1=\ldots=\eta_n = 0\}$
\end{definition}

We are now prepared to describe the oscillatory integrals.   A distribution
(generalized half-form) in $I^m(p,\Ls)$ can be written as a finite sum of
locally defined oscillatory integrals.  Specifically, given a
non-degenerate phase function $\phi$ parametrizing $\Ls$ locally,
we can write the distribution locally as
\be\label{oscint}
\int e^{i\phi(x,\tau,\eta)} a\Bigl(x,\tau,{\eta\over\sqrt\tau}\Bigr) d\tau
\> d\eta ,
\ee
where the amplitude $a(x,\tau, u)$  has the following properties (see
\S3 of \cite{BG} for the precise formulation of the estimates):
\begin{enumerate}
\item $a(x,\tau, u)$ is rapidly decreasing as a function of $u$.
\item $a(x,\tau,u)$ is cutoff to be zero near $\tau=0$.
\item For sufficiently large $\tau$, $a(x,\tau,u)$ admits an expansion of the
form
\be\label{aexp}
a(x,\tau,u) \sim \sum_{i = 0}^\infty \tau^{m_i} a_i(x,u),
\ee
where each $m_i$ is either integer or half-integer, with $m_0 = m-1/2$
and $m_i \to -\infty$.
\end{enumerate}
A change in the cutoff function used to enforce item 2 results only in a
smooth correction to (\ref{oscint}).  Because of this, the cutoff function is
generally suppressed from the notation.

Our next task is to actually construct a phase function that is linear
in $\tau$ and $\eta$.
Suppose that $\phi(x,\tau,\eta) = \tau f(x) + \sum_{j=1}^n\eta_j g_j(x)$.
We will hereafter adopt a vector notation
\be
\eta\cdot g := \sum_{j=1}^n\eta_j g_j(x), \qquad \eta\cdot dg :=
\sum_{j=1}^n \eta_j dg_j.
\ee
The critical set is $C_\phi = \{(x,\tau,\eta); f(x)
= g_1(x) = \ldots = g_n(x) = 0\}$, and the map $F$ is given by
$F(x,\tau,\eta) = (x;\>\tau \,df(x) + \eta\cdot dg(x))$.
In order for
$\phi$ to parametrize $\Ls$, we take the conic subset $B$ to be
$(\bbR_{+}\times\bbR^n)\mzs$, and choose
functions $f$ and $g_j$ satisfying two conditions.  We require that the
zero locus $\{f(x) = g_1(x) = \ldots = g_n(x) = 0\}$ define
$\Lambda$ in our local patch, and also that $df_x = \alpha_x$ for
$x\in\Lambda$ (locally).

By the Darboux theorem for contact manifolds, we can introduce local
coordinates $\{q_i, p_i, \theta\}$ on $P$ such that
\be
\alpha = \theta - p\cdot dq.
\ee
Because $\Lambda$ is Legendrian, by taking a small enough neighborhood
we can assume there exists a local generating function which gives the
relationship between $p$ and $q$ on $\Lambda$.  At least one of the
following two cases will occur:

\medskip\noi
{\it Case 1.}
There exists a function $h(q)$ such that $p_j = {\del h\over \del q_j}$ on
$\Lambda$.  In this case we take
\be
f = \theta - h, \qquad\qquad
g_j = p_j - {\del h\over \del q_j}.
\ee

\medskip\noi
{\it Case 2.}
There exists a function $h(p)$ such that $q_j = {\del h\over \del
p_j}$ on $\Lambda$.  We take
\be
f = \theta + h - p\cdot q, \qquad\qquad g_j = q_j - {\del h\over \del p_j}.
\ee
\medskip
Note that in either case $df = \alpha$ on $\Lambda$.

If we write $u\in J^m(P,\Lambda)$ as an integral of the form (\ref{oscint}),
the highest order term in the expansion (\ref{aexp}) is determined by the
symbol of $u$ computed in \S2.  The symbol map associating amplitudes
with symplectic spinors is given as follows.  Let $\pi$ be the projection
$P\times B \to P$.  The pull-back $\pi^*$ extends to a
morphism on half-forms once we fix the convention that $\pi^*\sqrt{dx} =
\sqrt{dx\> d\tau\> d\eta}$.   This map is an FIO with canonical
relation $\Gamma$ given by the conormal bundle of the graph $\pi$ in
$T^*(P\times P\times B)$.   We can parametrize
\be\label{gammapar}
\Gamma = \{(x,x,\tau,\eta; \xi,\xi,0,0)\}.
\ee
The symbol of $\pi^*$ is just the canonical half-form on $\Gamma$, which
in terms of these coordinates is just $\sqrt{dx\> d\xi\> d\tau\> d\eta}$.

Consider $d\phi$ as a map $P\times B \to
T^*(P\times B)$.  We have
\be
d\phi:(x,\tau,\eta) \mapsto (x,\tau,\eta; \tau df + \eta\cdot dg, fd\tau,
 g\cdot d\eta).
\ee
Let $\Sigma_\phi$ denote the image under $d\phi$ of the subspace $\eta_1
= \ldots = \eta_n = 0$,
\be
\Sigma_\phi = \{(x,\tau, 0; \tau df, f d\tau, \sum g_j d\eta_j)\}.
\ee
$\Sigma_\phi$ is an isotropic submanifold of $T^*(P\times B)$.
Furthermore, $\Gamma$ intersects $\Sigma_\phi$ transversally
and $\Gamma\circ\Sigma_\phi = \Ls$.

We define a symplectic spinor on $\Sigma_\phi$ by
\be\
\kappa = \sqrt{dx\> d\tau} \otimes \tau^{m-1/2} a_0(x, \eta)
\sqrt{d\eta},
\ee
where $a_0 (x, \eta)$ is the leading term of the expansion
(\ref{aexp}).
\begin{definition}
The symbol $\sigma(u)\in SS^m(\Ls)$ is the image of $\kappa$ under the
canonically defined map (see \cite{G2})
\be\label{ssmap}
SS^m(\Sigma_\phi) \to SS^m(\Gamma\circ\Sigma_\phi).
\ee
\end{definition}
The symbol map in this definition comes from the composition formula used
in Proposition \ref{legsym}.  We note that in \cite{BG}, there is an apparent
typo in the degree of homogeneity in $\tau$ of the amplitude $a_0$.

For the following Proposition, let $f$ and $g$ be chosen as above in
accordance with either Case 1 or Case 2.  For Case 1 let $H_{jk} = {\del^2
h\over \del q_j\del q_k}$, and for Case 2, $H_{jk} =
{\del^2 h\over \del p_j\del p_k}$.  In terms of the Darboux
coordinates, we write the metric as a matrix
\be
g = \pmatrix{A&B\cr B^t&D\cr},
\ee
with $A$ and $D$ symmetric.  The matrix of the symplectic form is
\be
\Omega = \pmatrix{0&I\cr -I&0\cr},
\ee
and the complex structure is given by $J = \Omega^t g$.  The requirement
that $J^2 = -I$ implies the following conditions:
\begin{eqnarray}
AD-B^2 & = & I,\nonumber\\
B^tD & = & DB,\label{gids}\\
AB^t & = & BA.\nonumber
\end{eqnarray}

\begin{proposition}\label{symprop}
Let $u\in J^m(P,\Lambda)$ with symbol $e\otimes\nu \in
H_\infty((\Ls)\prp/\Ls) \otimes \hffm \Ls$.  We can write $u$ locally as
an oscillatory integral of the form (\ref{oscint}) with
\be
a_0(x, \eta) = C_n \tilde\nu(x) \det M^{-1/2}
e^{-{1\over2}\eta^t M\inv \eta},
\ee
where $C_n$ depends only on the dimension, $\tilde\nu$ is an extension of
$\nu$ to be defined below, and
\be
M = \cases{(I+iB^t+iDH)\inv D & for Case 1,\cr
(I-iB-iAH)\inv A&for Case 2\cr}
\ee
(note that $M$ is symmetric in either case).
\end{proposition}
\begin{proof}
We need to compute the preimage of $e\otimes \nu$ under the symbol map
(\ref{ssmap}) associated to $\pi^*$.  The details of the map are given in
Proposition 6.5 of \cite{BG}.  This map can be broken into two parts: the map
of half-forms which takes $\tilde\nu$ to $\nu$ and the map of Schwarz
functions which takes the Gaussian above to $e$.  Fix a point $x =
(p,q,\theta) \in \Lambda$ and  $\tau \in \bbR_+$.   Let $W = T_{(x, \tau, 0;
\tau\alpha_x, 0, 0)}T^*(P\times G)$.  Recall that $G =
\bbR_+\times\bbR^n$
and  $\Sigma = T_{(x,\tau)}$.  Similarly let $\Gamma$ now denote the
tangent space to the $\Gamma$ defined above.

The half-form part of the map is particularly trivial in this case.  We
have an exact sequence
\be
0 \to \Gamma\circ\Sigma \to \Gamma\oplus \Sigma \to W \to 0,
\ee
which, together with the natural half-forms on $\Gamma$ and $W$,
furnishes an isomorphism
\be\label{sgsiso}
\hffm \Sigma \cong \hffm (\Gamma\circ \Sigma).
\ee
The map of half-forms reduces simply to this isomorphism,
so that $\tilde\nu$ can be any smooth function on $P$ such that the
isomorphism (\ref{sgsiso}) takes $\tilde\nu(x) \sqrt{dx\>d\tau}$ to $\nu$
at points of $\Ls$.

The non-trivial portion of the symbol map is really the isomorphism of the
Schwarz spaces $H_\infty(\Sigma\prp/\Sigma) \cong
H_\infty((\Gamma\circ\Sigma)\prp/(\Gamma\circ\Sigma))$, which
arises from a canonical symplectic isomorphism $\Sigma\prp/\Sigma \cong
(\Gamma\circ\Sigma)\prp/(\Gamma\circ\Sigma)$ (Proposition 6.4 of
\cite{BG}).  This map is given as follows.  Given $a\in
\Sigma\prp/\Sigma$, we choose $(b,c)\in \Gamma$ such that $c\in
\Sigma\prp$ and the image of $c$ in $\Sigma\prp/\Sigma$ is $a$.  Then
$b\in \Gamma\circ\Sigma\prp = (\Gamma\circ\Sigma)\prp$ and the
association $a\to b$ descends to an isomorphism when we mod out by
$\Sigma$.   Because $e$ was defined through the identification of
$(\Gamma\circ\Sigma)\prp/(\Gamma\circ\Sigma)$ with the horizontal
subspace of $T_xP$, we will construct the map directly to this horizontal
subspace.

We break the problem up into the two cases described above.
Assume first that we are in Case 1, where $p = {\del h\over \del q}$ on
$\Lambda$ and $f = \theta - h$.  Define $H_{jk} = {\del^2h\over\del
q_j\del q_k}$.
In terms of the basis $\{{\del \over \del q_j}, {\del\over\del p_j},
{\del\over\del\theta}, \ldots\}$, a straightforward computation gives
\be
\Sigma = \{(v, w, t, r, 0; -rp -Hv, 0, r, t-p\cdot v, w - Hv)\}
\ee
(where $v$ and $w$ are $n$-vectors and $t$ and $r$ are real numbers).
{}From this we compute that
\be
\Sigma\prp = \{(v, w, t,r,\beta; v-rp-w-H\beta, \beta, r, t - p\cdot v,
\gamma)\}
\ee
Define $\psi:\Sigma \to \bbR^n\times\bbR^n$ by
\be
\psi: (v, w, t,r,\beta; v-rp-w-H\beta, \beta, r, t - p\cdot v, \gamma)
\mapsto (\beta, \gamma - w + Hv).
\ee
The kernel of this map is $\Sigma$, so it descends to an isomorphism
$\Sigma\prp/\Sigma\to \bbR^n\oplus\bbR^n$ (with the natural
symplectic structure on the latter).   We will henceforth
identify these spaces through this isomorphism, giving
$\Sigma\prp/\Sigma$
the coordinates $(\beta, \sigma)$.

To $a = (\beta, \sigma) \in \Sigma\prp/\Sigma$ we associate the vector $c =
(0, -\sigma, 0, 0,\beta;  \sigma - H\beta, \beta, 0, 0, 0) \in \Sigma\prp$,
so that $\psi(c) = a$.   In view of (\ref{gammapar}), for $(b,c)$ to be in
$\Gamma$ we must take $b = (0,-\sigma,0; \sigma-H\beta,\beta, 0) \in
\Gamma\circ\Sigma\prp$.   In the notation of \S2, we have a
decomposition $V = T_{(x, \tau\alpha_x)}(T^*P) = Z\oplus Z\prp$, and
terms of the Darboux coordinates:
\be
Z = \{(v,w,t;-rp-w,0,r)\},
\ee
and
\be
Z\prp = \{(v,w,p\cdot v;-rp-w,0,r)\}.
\ee
Recall that the linearization of the symplectic normal to $\Ls$ (here written
$(\Gamma\circ\Sigma)\prp/(\Gamma\circ\Sigma)$) was isomorphic to
the symplectic vector space $Z\prp$.  Thus to complete the map we need only
project $b\in \Gamma\circ\Sigma\prp$ into $Z\prp$.   In order to
pull-back the Gaussian $e$ we then identify
$Z\prp$ with the horizontal subspace of $T_xP$ by  $(v,w,p\cdot v;-rp-w,0,r)
\to (v,w)$.   The result is that the symplectic isomorphism from
$\Sigma\prp/\Sigma$ to $Z\prp$ can be written
\be\label{bssymp}
(\beta, \sigma) \mapsto (-\beta, -\sigma-H\beta),
\ee
where the symplectic structure on both sides is the natural structure on
$\bbR^n\oplus\bbR^n$.

Consider how $e$ was constructed (Proposition 4.2 of
\cite{BG}).  If the set of complex vectors $\{x_j + iy_j\}$ gives a basis
for the (1,0)-subspace
of the complexified horizontal tangent space to $P$ at $x$, then $e$ is
defined as a solution (of norm one) to the equations
$(x_j + iy_j)e = 0$, where $y_j$ acts as $-i{\del\over\del x_j}$.
In the present case, given that
\be
J = \pmatrix{-B^t&-D\cr A&B\cr},
\ee
the (1,0)-subspace is spanned by vectors of the form $\{v - iB^tv + iDw\}$.
Composing with the inverse of the symplectic map (\ref{bssymp}), we see
that $e(\beta)$ should satisfy the differential equation
\be
\biggl[ \beta - iB^t\beta + iD\Bigl(-i{\del\over\del\beta} +
\beta\Bigr)\biggr] e(\beta) = 0 .
\ee
The solution (up to a constant depending only on the dimension) is
\be
e(\beta) =  \det M^{-1/2} e^{-{1\over2}\beta^t M\inv\beta}
\sqrt{d\beta},
\ee
where $M = (I+iB^t+iDH)\inv D$.  The factor $\det M^{-1/2}$ appears because
$e$ transforms as a half-form under symplectic transformations.
This completes the proof for Case 1.

The proof is quite similar for Case 2.  Here we have $q_j = {\del h\over \del
p_j}$ as the defining relation of $\Lambda$.  Taking $H_{jk} = {\del^2
h\over \del p_j\del p_k}$, the symplectic map (\ref{bssymp}) turns out to be
\be
(\beta, \sigma) \mapsto (-\sigma+H\beta, \beta).
\ee
This leads to the Gaussian given above.
\end{proof}

\begin{theorem}\label{localuk}
Let $u\in J^m(P, \Lambda)$ with symbol $\nu$.  Choose $f$ and $g$ locally
as above, and define $f_0(p,q) := \theta - f(p,q,\theta)$.  For sufficiently
large $k$ the $k$-th isotype of $u$ under the $S^1$ action has the local
representation
\be
u_k =  C_{m,n} k^{m+(n-1)/2} \tilde\nu(p,q,f_0)
e^{ikf - {k\over 2}g^tMg} + O(k^{(m+n/2-1)})
\ee
(in the sup norm topology), where $C_{m,n}$ is a constant depending only on
$m$ and $n$.  (In fact, there is full asymptotic expansion in decreasing
half-integer powers of $k$.)
\end{theorem}
\begin{proof}
We start with the local oscillatory integral representation
\be\label{uosc}
u(p,q,\theta) = \int e^{i\tau f + i\eta\cdot g}
\>a\Bigl(p,q,\theta, \tau, {\eta\over\sqrt\tau}\Bigr) d\tau \>d\eta
\ee
with
\be
a(p,q,\theta,\tau,u) \sim \sum_{j=0}^\infty \tau^{m_j} a_j(p,q,\theta, u).
\ee
To pick off the $k$-th isotype, we project onto the $e^{ik\theta}$
component by integrating the above expression against $e^{-ik\theta}$ for
$0\le\theta\le 2\pi$.  The
expression (\ref{uosc}) is cutoff in $\theta$ so we may in fact extend the
integration limits to infinity.

We will consider one term in the expansion at a time.  Let
\be
W_k(p,q,\theta) =  e^{ik\theta} \int e^{-ik\theta'}
e^{i\tau f(p,q,\theta')} \tau^{l}
a\Bigl(p,q,\theta',{\eta\over\sqrt\tau}\Bigr) d\tau \>d\theta'.
\ee
Rescaling $\tau \to k\tau$ yields
\be
W_k(p,q,\theta) =  k^{l+1} e^{ik\theta} \int e^{-ik\theta'}
e^{ik(-\theta'+\tau\theta' - \tau f_0)}
\tau^{l}
a\Bigl(p,q,\theta',{\eta\over\sqrt{k\tau}}\Bigr) d\tau\>d\theta'
\ee
This expression can be estimated for large $k$ by stationary
phase.  The only stationary point occurs at $\tau =1$, $\theta'=f_0(p,q)$, so
we obtain the estimate
\begin{eqnarray}
\lefteqn{\left| W_k(p,q,\theta) - 2\pi k^l e^{ikf} \Bigl[ 1 + k\inv
L_{\tau,\theta'} \tau^l
a\Bigl(p,q,\theta',{\eta\over\sqrt{k\tau}}\Bigr)\Bigr] \right|} \\
& & \le C k^{l-1} \sum_{|\alpha| \le 4}  \sup \Bigl| D^{\alpha} \tau^l
a\Bigl(p,q,\theta',{\eta\over\sqrt{k\tau}}\Bigr) \Bigr|,\nonumber
\end{eqnarray}
where $L_{\tau,\theta'}$ is a second order differential operator in $\tau$
and $\theta'$, evaluated at
the critical point, and $D$ represents only derivatives with respect to
$\tau$ and $\theta'$.   Note that the sup is finite if and only if $l \le 0$.
By applying successive integrations by parts in the original expression, we
may assume that this is the case.

Derivatives of $a$ with respect to $\tau$ bring out a factor of
$k^{-1/2}$, and derivatives with respect to $\theta'$ have no effect in terms
of $k$.   The first correction and the error term are thus both well-behaved in
terms of $k$.   So in terms of the sup norm we have
\be
W_k(p,q,\theta) = 2\pi k^l e^{ikf} + O(k^{l-1}).
\ee

Applying this result to $u_k$, we obtain
\be
u_k(p,q,\theta) =  2\pi k^{m-1/2} e^{ik(\theta - f_0)} \int
e^{i\eta\cdot g}  a_0\Bigl(p,q,f_0,{\eta\over\sqrt{k}}\Bigr) d\eta
+ O(k^{m-1}).
\ee
{}From Proposition \ref{symprop}, we see that the remaining $\eta$
integration is just the Fourier transform of a Gaussian:
\be
 \int e^{i\eta\cdot g} e^{-{1\over2k}\eta^t M\inv \eta} d\eta =
C k^{n/2} \det M^{1/2} e^{-{k\over2} g^tMg}.
\ee
\end{proof}

\subsection{Proof of the estimates}

For this subsection, we assume that $u = \Pi(\delta_{\nu_1}) \in
J^{1/2}(P,\Lambda_1)$ and $v = \Pi(\delta_{\nu_2}) \in
J^{1/2}(P,\Lambda_2)$.   Consider the inner product $\ip{u_k, v_k}$.
Because of the reproducing property of the kernel of $\Pi$ and the fact that
the states come from delta functions, $\ip{u_k, v_k}$
can be written as the integral of $u_k$ (represented as a function on $P$)
over $\Lambda_2$ (with a measure determined by $\nu_2$).

Because of this fact, the following proposition (a restatement of Corollary
\ref{maincor}) follows from Theorem \ref{localuk}.
\begin{proposition}\label{nprop}
Let $u = \Pi(\delta_\nu) \in J^{1/2}(P,\Lambda)$, and suppose that
$\Lambda$ is a $k_0$-fold covering of $\pi(\Lambda)$, with $\nu$
invariant under the action of the covering group.   Then as $k\to\infty$
\be
\ip{u_k, u_k} \sim k_0\left({k\over\pi}\right)^{n/2} \int_\Lambda
|\nu|^2,
\ee
for $k$ a sufficiently large multiple of $k_0$.
\end{proposition}
\begin{proof}
{}From Theorem \ref{localuk} and the reproducing property we find
\be
\ip{u_k, u_k} \sim C \sum_{j=1}^{k_0} e^{2\pi i{kj\over k_0}}
k^{n/2} \int_\Lambda |\nu|^2.
\ee
The sum is zero unless $k_0$ divides $k$, in which case it yields the
factor $k_0$.

Since the constant out front is universal, it may be computed in a
particular example.  This is easily done for the Bargmann space $\bbC^n$,
where the kernel of $\Pi$ can be written explicitly.
\end{proof}

\bigskip
Let $\Lambda_1$ and $\Lambda_2$ be two distinct intersecting Legendrian
submanifolds.
The following lemmas are a prelude to taking the stationary phase
approximation of the inner product of states defined on $\Lambda_1$ and
$\Lambda_2$.  Let $f$ and $g_j$ be chosen to parametrize $\Lambda_1$ as in
the last subsection (according to whether $\Lambda_1$ satisfies Case 1 or
2).  Recall that the highest order term in $u_k$ involved the phase function
$\psi = f + {i\over2}g^tMg$, with $M$ determined by $f$ and $g$.
Choose a set of local parameters $\{t_1,\ldots, t_n\}$ to describe
$\Lambda_2$, so that $\{{\del\over\del t_1},\ldots,{\del\over\del t_n}\}$
gives locally an orthonormal frame for $T\Lambda_2$.
We will use primes to denote derivatives with respect to these parameters.

\begin{lemma}\label{stpts}
The stationary points of $\psi$, with respect to the parametrization
described above, correspond precisely with points of $\calP$.
\end{lemma}
\begin{proof}
The proof is similar for either case, so we assume Case 1.  Then
\be
\psi(p,q,\theta) = \theta -  h(q) + {i\over2} \Bigl(p - {\del h\over\del
q}\Bigr) M \Bigl(p - {\del h\over \del q}\Bigr),
\ee
with derivative
\be
\psi'(p,q,\theta) = \theta' -  {\del h\over \del q}q' +
i\Bigl(p - {\del h\over\del q}\Bigr) M (p' - Hq')
+ {i\over2} \Bigl(p - {\del h\over\del q}\Bigr)
M' \Bigl(p - {\del h\over \del q}\Bigr).
\ee
Since the horizontality of $\Lambda_2$ implies that $\theta' = pq'$, we see
immediately that $p = {\del h\over\del q}$ implies $\psi' = 0$.  This is
the case whenever there exists an $\omega\in S^1$ and $x\in\Lambda_1$
such that $x\cdot\omega \in\Lambda_2$.

It follows directly from Proposition \ref{Upsmooth} that there are no other
possible stationary points.
\end{proof}

\begin{lemma}\label{hess}
The Hessian of $\psi =  f + {i\over 2}g^tMg$ at a
stationary point $x$
\be
\psi'' = \xi_2^t \Omega^t \xi_1 (\xi_1^t g\xi_1)\inv
\xi_1^t (g+i\Omega) \xi_2,
\ee
where $\xi_2 = \pmatrix{q'\cr p'\cr}$ and $\xi_1 = \pmatrix{I\cr H\cr}$.
\end{lemma}
\begin{proof}
The Hessian of $\psi$ is
\be
\psi'' = p'^tq' - q'^tHq' + i(p'^t-q'^tH)M(p'-Hq'),
\ee
where we have used the fact that $\theta' = pq'$ because $\Lambda_2$ is
horizontal.  Recalling the definition of $M$, we can write this as
\begin{eqnarray}
\psi'' & = & (p' - Hq')^t \Bigl[q' + i(I+iB^t+iDH)\inv D(p'-Hq') \Bigr]\\
& = & (p'-Hq')^t (I+iB^t+iDH)\inv [q'+iB^tq' + iDp'].\nonumber
\end{eqnarray}
We insert the matrix $(I-iB^t-iDH)$ and its inverse and use the identities
(\ref{gids}) to obtain
\begin{eqnarray}
\psi'' & = & (p'-Hq')^t (A+BH+HB^t +HDH)\inv \\
& & \times \Bigl[(A+HB^t)q' + (B+HD)p' + i(p'-Hq') \Bigr].\nonumber
\end{eqnarray}
Note that this is exactly the expression given above.
\end{proof}

\begin{proposition}\label{ipprop}
The inner product $\ip{u_k, v_k}$ has an asymptotic expansion whose terms
correspond to elements of $\Phi(\calP)$.  The leading contribution from a
particular $\omega\in \Phi(\calP)$ is
\be
(2i)^{(n-d)/2} \left({k\over\pi}\right)^{d/2}
\omega^k \int_{\Lambda_1\cdot\omega \cap \Lambda_2}
\det\{\Lambda_1\cdot\omega, \Lambda_2\}^{-1/2}
\nu_1 \hatch \ol{\nu_2},
\ee
where $d$ is the dimension of $\Phi\inv(\omega)$.
\end{proposition}
\begin{proof}
We start with the representation of Theorem \ref{localuk}.
Choose the parametrization of $\Lambda_2$ so that the first $d$ variables
parametrize $\Lambda_1\cdot\omega \cap\Lambda_2$.  The method of
proof is to apply stationary phase to the integral over the $n-d$ transverse
parameters of $\Lambda_2$.   The integral over the remaining $d$
variables survives in the final expression.
For notational clarity, consider only the case where
$\Lambda_1\cdot\omega$ and $\Lambda_2$ intersect transversally
($d=0$).

{}From Theorem \ref{localuk}, the highest order term contribution to the
inner product is
\be
\ip{u_k, v_k} = \left({k\over\pi}\right)^{n/2} \int
\ol{\nu_2}(p,q,\theta) \tilde\nu_1(p,q,f_0) \>e^{ikf - {k\over2}g^tMg} d^nt
+ O(k^{(n-1)/2}),
\ee
where we have filled in the constant based on the computation in
Proposition \ref{nprop}.
According to Lemma \ref{stpts}, when we apply stationary phase to this
integral, we obtain a term for each component of $\calP$, i.e. for each
point in $\Phi(\calP)$.

In the transverse case, at the point $x\in
\Lambda_1\cdot\omega\cap\Lambda_2$, the stationary phase lemma
yields the term
\be\label{trcase}
\left({k\over\pi}\right)^{n/2}  \nu_1(x\cdot\omega) \ol{\nu_2}(x)
\omega^k
\left({2\pi i\over k}\right)^{n/2} (\det\psi'')^{-1/2}.
\ee

We can reinterpret Lemma \ref{hess} in the following way.  Given an
orthonormal basis $\{e_i\}$ for $T_x\Lambda_1$ and $\{f_i\}$ for
$\Lambda_2$,
\be
\det\psi'' = \det \{\omega(e_i,f_j)\} \det \{h(e_i,f_j)\},
\ee
where $\omega$ is the symplectic form  and $h$ the hermitian form.  The
first term on the right-hand side is (when raised to the $-1/2$ power) the
factor which appears
in the construction of $\nu_1\hatch\ol{\nu_2}$ when we divide out by the
square root of the Liouville form.   The second term is the function
$\det \{\Lambda_1\cdot\omega, \Lambda_2\}$, as defined in \S3.1.

In general, the stationary phase approximation is done over $n-d$
variables, so the last factors in (\ref{trcase}) are $(2\pi i/ k)^{(n-d)/2}
(\det \psi'')^{-1/2}$, with the determinant taken over the transverse
variables.  It is straightforward to see the this determinant yields again
the intersection of the half-forms with the same unitary factor.
\end{proof}

\bigskip
To conclude this section, we note that Theorem \ref{mainthm} follows
directly from Propositons \ref{nprop} and \ref{ipprop}.  The insertion of
the Toeplitz operator $T_A$ in the inner product is an essentially trivial
modification.

\section{Bohr-Sommerfeld curves and Poincar\'e series}
\newcommand{\BSk}{\mbox{BS}_k}
\newcommand{\BSko}{\mbox{BS}_{k_0}}
\newcommand{\im}{\mathop{\mbox{Im}}}
\newcommand{\hil}{{\calS}}
\newcommand{\hilk}{{\calS_k}}
\renewcommand{\inv}{^{-1}}
\newcommand{\tildg}{{\tilde\gamma}}
\renewcommand{\tildx}{{\tilde\xi}}
\newcommand{\sgn}{\mathop{\mbox{sgn}}}
\newcommand{\pmid}{\{\pm\hbox{\it id}\}}

In this Section we examine in the case where $X$ is the
quotient of the upper half plane by a Fuchsian group of the first kind.  The
natural quantizing line bundle $L$ is simply the holomorphic tangent
bundle.  We will perform the general constructions outlined in Section I
quite explicitly for this case.  In particular, we compute the states
associated to Bohr-Sommerfeld curves given by hypercycles, horocycles, or
circles in $H$ and show that these correspond to well-known Poincar\'e
series.

\subsection{Bohr-Sommerfeld curves}
Let $H$ be the upper-half plane $\{z\in\bbC: \im z>0\}$, and let $SH$
denote the unit circle bundle of the holomorphic cotangent bundle of $H$,
\be
SH = \{(z,\zeta)\in H\times\bbC: |\zeta| = \im z\}.
\ee
The group $G = \SL(2,\bbR)$ acts on $SH$ by fractional linear
transformations.  In fact, there is a homeomorphism $SH \cong G/\pmid$,
such that the action of $G$ on $SH$ corresponds to the left
action of $G$ on $G/\pmid$.  Explicitly, for
\be
g = \pmatrix{a&b\cr c&d\cr}\in G,
\ee
we have
\be
g\cdot(z,\zeta) = (g\cdot z, j(g,z)^{-2}\zeta),
\ee
where
\be
g\cdot z = {az+b\over cz+d}
\ee
and $j(g,z) = cz+d$.   $G$ is represented on the space of functions on $SH$ by
\be
(g\cdot F)(z,\zeta) = F(g\inv\cdot(z,\zeta)).
\ee
The contact form $\alpha$ is given by
\be
\alpha = d\phi - {dx\over y},
\ee
where $z = x+iy$, and $\zeta = ye^{i\phi}$.  The volume form $dV = (2\pi)\inv
\alpha \wedge d\alpha$ is
the $G$-invariant volume form on $SH$:
\be
dV = {dx\>dy\over y^2} \>{d\phi\over 2\pi}.
\ee

\medskip
The connection on $SH$ corresponding to $\alpha$ is naturally defined
as follows.   Letting $I$ denote the point $(i,1)\in SH$, we identify
$G/\pmid$ with $SH$ by the map $g\mapsto g\cdot I$.
We thus have $T_I(SH) \cong
\mbox{sl}(2,\bbR) = \frakt\oplus\frakp$.  Since we also know that
$T_iH \cong \frakp$, we can define a left-invariant connection simply by
declaring $\frakp$ to be the horizontal space at $I\in SH$.
Using the identification $T_{g\cdot I}SH \cong T_gG$, we see that the
horizontal tangent space of $T_{g\cdot I}SH$ is $g\cdot\frakp$.   Therefore, if
$g(t):\bbR\to G$ is a smooth curve, $g(t)\cdot I$ will be horizontal iff
\be
g(t)\inv\cdot \dot g(t) \in\frakp,
\ee
i.e., $g(t)\inv\cdot \dot g(t)$ must be traceless and symmetric.

\medskip
Recall that $SH$ is the boundary of a strictly pseudoconvex domain (the unit
disk bundle over $H$).  Let $\hil(H) \subset L^2(SH)$ denote the Hardy
space of boundary values of holomorphic functions on the unit disk bundle.
The $k$-th isotype of $\hil(H)$ under the action of $S^1$ is the space of
holomorphic $k$-differentials on $H$, which we denote by $\hilk(H)$.  This
consists of functions $F:SH\to \bbC$ of the form $F(z,\zeta) = \zeta^k f(z)$,
where $f:H\to\bbC$ is holomorphic.  In other words,
\be
\hilk(H) = \left\{ \zeta^k f(z): \int_H |f(z)|^2\> y^{2k}\> {dx\>dy \over y^2}
<\infty\right\}.
\ee

\medskip
Let $\Gamma$ be a discrete subgroup of $G$ such that $X = \Gamma\bs H$
has finite volume, i.e., a Fuchsian group of the first kind.   The unit circle
bundle of the holomorphic cotangent bundle of $X$, denoted by $SX$, is
again the boundary of a strictly  pseudoconvex domain, and we let $\hil(X)$
denote the Hardy space for this  domain.  As above, $\hilk(X)$ denotes the
space of holomorphic  $k$-differentials on $X$.  Define the orthogonal
projections    \be
\Pi: L^2(SX) \to \hil(X)\quad\hbox{and}\quad\Pi_k:L^2(SX) \to \hilk(X).
\ee
If $F$ is a function on $SH$ which is invariant under the action of $G$, then
$F$ corresponds to a function on $SX$.  Thus we can identify $\hilk(X)$ with
the space of cusp forms $S_{2k}(\Gamma)$:
\be
\hilk(X) = S_{2k}(\Gamma) = \left\{f(z): f(g\cdot z) = f(z) j(g,z)^{2k},
\; \int_\calF |f(z)|^2 y^{2k} {dx\>dy\over y^2} <\infty\right\},
\ee
where $\calF$ is a fundamental domain for $\Gamma$.

\medskip
In what follows,
by a smooth closed curve with domain $[0, T]$ we mean the restriction
to $[0,T]$ of a smooth $T$-periodic map with domain $\bbR$.  Generally, we
will think of a closed curve $\gamma$ on $X$ as the projection to $X$ of a
curve $\gamma:\bbR\to H$ such that the points $\gamma(t)$ and
$\gamma(t+T)$ are related by an element of $\Gamma$.

\begin{definition}\label{BSK}
Let $k$ be a positive integer.
A parametrized smooth closed curve, $\gamma : [0,T]\to X$ is said to satisfy
the Bohr-Sommerfeld condition of order $k$, or $\BSk$ for short, iff
its holonomy in $SX$ is an $k$-th root of unit.
\end{definition}
\noi
Note that the $\BSk$ property is invariant under reparametrizations.  Also
note that a curve which satisfies $\BSk$ satisfies the BS condition for any
integer multiple of $k$.

\medskip
To any curve satisfying $\BSko$ we now describe how to associate a vector in
$\hilk(X)$, where $k$ is a multiple of $k_0$.

\begin{definition}\label{STATE}
Assume $\gamma$ satisfies $\BSko$, and let $\tilde{\gamma}$ be its
horizontal lift as in definition (\ref{BSK}).  If $\delta_{\tildg}$ denotes the
delta function integrated along $\tildg$ using the parametrization, for $k$ a
multiple of $k_0$ we define
\be\label{a.3}
\ket{\gamma ,k}\,=\, \Pi_k\,(\,\delta_{\tildg}\,)\,.
\ee
\end{definition}

\noi{\bf Remark.}
If $k$ is not a multiple of $k_0$ then the projection in (\ref{a.3}) clearly
zero gives zero.
The definition of $\ket{\gamma,k}$ depends on the choice
of the horizontal lift, $\tildg$, but it's easy to see that
changing the horizontal lift amounts to multiplying the state by a
complex number of modulus one.

\medskip
\begin{lemma}
Two $\BSk$ curves defined as above are immersed Lagrangian
submanifolds
of $X$ satisfying the cleanness assumption of \S3.1 provided they have no
common tangents.
\end{lemma}

\noi
Note that in particular any pair of geodesic $\BSk$ curves satisfy the
assumption (including a geodesic with itself).  We will see
below that all geodesic curves are BS$_1$.

In order to apply the theory developed in \S3, $X$ must be a compact
manifold, i.e. a Riemann surface.  In addition, $SX$ must be given a
metalinear structure.  To do this, note that $SX$ is naturally identified
with $G/\Gamma$, so that $T(SX) \cong P\times \lig$, where $\lig$ is the
Lie algebra of $G$.  Thus $SX$ inherits a metalinear structure from the
metalinear structure on the vector space $\lig$.
{}From this, we obtain metalinear structures on the $\BSk$ curves.  For
our purposes here these structures will be invisible, since by
Definition \ref{STATE} we will deal
only with half-forms defined through parametrizations.

{}From Theorem \ref{mainthm} and Corollary \ref{maincor} we
obtain the following result.
\begin{theorem}\label{rsmnthm}
Let $X$ be a Riemann surface and $\gamma$ a $\BSko$ curve with no
self-tangents, parametrized by arclength.  For $k$ a
sufficiently large multiple of
$k_0$ we have
\be
\braket{\gamma,k}{\gamma,k} = \left({k\over\pi}\right)^{1/2} k_0^2
T + O(1).
\ee
Furthermore, if $\gamma_1$ and $\gamma_2$ are two distinct intersecting
$\BSko$ curves with no common tangents, then for $k$ a sufficiently large
multiple of $k_0$,
\be
\braket{\gamma_1,k}{\gamma_2,k} = 2^{1/2} k_0^2
\sum_{p \in \gamma_1 \cap \gamma_2} {\omega_p^k e^{-i(\vartheta_p/2 -
\pi/4)}\over\sqrt{\sin \vartheta_p}} + O(k^{-1/2}),
\ee
where $\vartheta_p$ is the angle from $\gamma_1$ to $\gamma_2$ at $p$,
and $\omega_p\in S^1$ is determined by the condition that $\tildg_1\cdot
\omega_p$ intersects $\tildg_2$ over the point $p$.
\end{theorem}

\subsection{Relative Poincar\'e series}

We seek to write out the state $\ket{\gamma,k}$ explicitly as a function on
$SH$ that is invariant under $\Gamma$. Let
$\psi_{(w,\eta)}$ denote the coherent state in $\hilk(H)$ associated to the
point $(w,\eta)\in SH$, i.e., the function on $SH$ which is the orthogonal
projection of the delta function at $(w,\eta)$ into
$\hilk(H)$.  By definition, the coherent states are equivariant under the
action of $G$,
\be\label{psiequi}
g\cdot \psi_{(w,\eta)} = \psi_{g\cdot(w,\eta)}.
\ee
To obtain coherent states in $\hilk(X)$, we average over the action of $G$.  It
follows from a theorem of Katok \cite{Ka} that for any function
$F\in\hilk(H)$,
\be\label{pseries}
\sum_{g\in\Gamma} g\cdot F \in \hilk(X),
\ee
where the convergence is absolute and uniform on compact sets.
The coherent state in $\hilk(X)$ associated to an equivalence class
$[(w,\eta)]\in SX \cong \Gamma\bs SH$ is thus
\be
\Psi_{[(w,\eta)]} = \sum_{g\in\Gamma} g\cdot \psi_{(w,\eta)},
\ee
Because of the equivariance (\ref{psiequi}), the sum on the right depends
only on the class $[(w,\eta)]$.

\medskip
The following proposition realizes our states $\ket{\gamma,k}$ as relative
Poincar\'e series for functions given by integrals over coherent states.
The proof is clear from the absolute convergence of (\ref{pseries}).
\begin{proposition}
Let $\gamma$ be a $\BSk$ curve on $X$ and let $\tildg$ be a horizontal lift
of $\gamma$ to $SX$.  Then the state $\ket{\gamma,k}\in\hilk(X)$
corresponds to the
function
\be\label{gammasum}
\Phi_{\gamma}(z,\zeta) = \sum_{g\in \Gamma} g\cdot F(z,\zeta),
\ee
where $F(z,\zeta)$ is given by
\be
F(z,\zeta) = \int_0^T \psi_{\tildg(t)}(z,\zeta) dt.
\ee
\end{proposition}

\noi
We can improve upon the realization given above if $\gamma$ is not closed
as a curve on $H$, using the Rankin-Selberg technique.

\begin{proposition}\label{RPS}
Suppose that $\gamma_0\in\Gamma$ is not elliptic, and let $\gamma$ be a
$\BSk$ curve defined as a map $\gamma:\bbR\to H$ such that
$\gamma_0\cdot\gamma(t) = \gamma(t+T)$.  Then the state
$\ket{\gamma,k}$ corresponds to the function
\be
\Phi_{\gamma}(z,\zeta) = \sum_{g\in \Gamma_0\bs\Gamma}
\int_{-\infty}^\infty \psi_{\tildg(t)}(g\cdot(z,\zeta)) \>dt,
\ee
where $\Gamma_0$ is the cyclic group $\langle\gamma_0\rangle$ and
$\tildg$ is a horizontal lift of $\gamma$ to $SH$.
\end{proposition}\begin{proof}
First of all, note that since the connection is left-invariant, the horizontal
lift
$\tildg$ also satisfies $\gamma_0\cdot \tildg(t) = \tildg(t+T)$.
We break the sum over $\Gamma$ in (\ref{gammasum}) up into a sum
over cosets of $\Gamma_0$:
\be
\Phi_{\gamma}(z,\zeta)
= \sum_{g\in \Gamma/\Gamma_0} \sum_{n=-\infty}^\infty
\int_0^T (g\gamma_0^n)\cdot\psi_{\tildg(t)}(z,\zeta) dt.
\ee
Now, since
\be
\gamma_0^n\cdot \psi_{\tildg(t)} = \psi_{\tildg(t+nT)},
\ee
we can reduce the sum in the above integral to
\be
\Phi_{\gamma}(z,\zeta) = \sum_{g\in \Gamma/\Gamma_0}
\int_{-\infty}^\infty g\cdot\psi_{\tildg(t)}(z,\zeta) dt.
\ee
To complete the proof, we note that $g\in \Gamma/\Gamma_0$ implies
$g\inv\in \Gamma_0\bs\Gamma$.
\end{proof}

\medskip
In order to realize the relative Poincar\'e series given in these propositions
more concretely we need to compute explicitly the coherent state
$\psi_{(w,\eta)}$.

\begin{lemma}
The orthogonal projection of the delta function at $(w,\eta)\in SH$ into
$\hilk(H)$ is the function
\be
\psi_{(w,\eta)}(z,\zeta) = A_k {\zeta^k \bar\eta^k\over (z-\bar w)^{2k}},
\ee
where
\be
A_k = (-1)^k 2^{2k-2}{2k-1\over \pi} .
\ee
\end{lemma}
\begin{proof}
The fact that $\psi_{(w,\eta)} = \Pi_k (\delta_{(w,\eta)})$ is equivalent to
the reproducing property:
\be
F(w,\eta) = \int_{SH} \ol{\psi_{(w,\eta)}(z,\zeta)} F(z,\zeta)\> dV,
\ee
for all $F\in\hilk(H)$.   Given any orthonormal basis $\{F_{l,k}\}$ for
$\hilk(H)$, we can write the reproducing kernel as the series
\be
\psi_{(w,\eta)}(z,\zeta) =  \sum_{l} \ol{F_{l,k}(w,\eta)} F_{l,k}(z,\zeta),
\ee
which converges absolutely and uniformly on compact sets.
Using the well-known orthonormal basis
\be\label{onbasis}
F_{l,k}(z,\zeta) = 2^{2k-1} \left[ {1\over \pi} {(2k+l-1)!\over (2k-2)! l!}
\right]^{1/2} \>\zeta^k {(z-i)^l\over (z+i)^{l+2k}},
\ee
we obtain the result given above.
\end{proof}

\subsection{Hypercycles and geodesics}

We consider now the application of Proposition \ref{RPS} to the special case
when $\gamma_0$ is hyperbolic and the associated curve $\gamma$ is a
hypercycle in $H$.   We first consider the question of when a hypercycle
corresponds to a $\BSk$ curve on $X$.

\begin{proposition}\label{bskhyp}
Let $\gamma_0\in\Gamma$ be hyperbolic, and suppose that $\gamma$ is a
hypercycle $\bbR\to H$ such that $\gamma_0\cdot\gamma(t) =
\gamma(t+T)$.
Let $\tau$ denote the cotangent of the angle from the real axis to $\gamma$
at the $-\infty$ limit point of $\gamma$.  Then $\gamma$ is $\BSk$ as a
curve on
$X$ if and only if
\be\label{tauj}
\tau = {2\pi j\over kT}
\ee
for some $j\in\bbZ$.  In particular, all geodesics (for which $\tau=0$)
are $\BSk$ curves for any value of $k$.
\end{proposition}
\begin{proof}
It suffices to consider the case where $\xi(t) = e^t(\tau+i)$,
since the $\BSk$ property is equivariant.
The horizontal lifting of $\xi$ to a curve on $G$ is
\be
g(t) = \pmatrix{e^{t/2}&\tau e^{t/2}\cr 0&e^{-t/2}\cr}
\pmatrix{\cos t\tau/2&-\sin t\tau/2\cr
\sin t\tau/2& \cos t\tau/2\cr}.
\ee
On $SH$ this corresponds to
\be\label{tildx}
\tildx(t) = (e^t(i+\tau), e^{t(1-i\tau)})
\ee
The $\BSk$ condition requires that $(e^{-iT\tau})^k=1$, which implies that
$kT\tau = 2\pi j$ for some $j\in\bbZ$.
\end{proof}

For the remainder of this subsection we will assume that $\gamma$ is a
$\BSko$ curve such that
\be
\tau = {2\pi j\over k_0T},
\ee
for some $j\in\bbZ$.
To compute the state associated to a general hypercycle, we first consider the
hypercycle which connects the origin to the point at infinity, and then use the
equivariance of the coherent states.

\begin{lemma}
Let $\xi$ be the hypercycle $\xi(t) = e^t(\tau+i)$ in $H$, with the lifting
$\tildx$ defined as in (\ref{tildx}).  Then
\be
\int_{-\infty}^\infty \psi_{\tildx(t)}(z,\zeta)\>dt = B_{k,\tau}
\zeta^k z^{ik\tau-k},
\ee
where
\be
B_{k,0} = (-i)^k {(k-1)!^2\over (2k-1)!},
\ee
and for $\tau\ne 0$,
\be
B_{k,\tau} = {2\pi i\over 1- e^{-2\pi k\tau}} (\tau-i)^{-k-i\tau}
{\Gamma(ik\tau+k)\over (2k-1)! \Gamma(ik\tau-k+1)}.
\ee
\end{lemma}\begin{proof}
We integrate the coherent states along the curve $\tildx$:
\be
\int_{-\infty}^\infty \psi_{\tildx(t)} (z,\zeta) dt =
\int_{-\infty}^\infty {\zeta^k e^{kt(1+i\tau)} \over (z -
e^t(-i+\tau))^{2k}} dt
\ee
The results are obtained by substituting $u = e^t$ and performing a contour
integration.
\end{proof}

Given a hypercycle $\gamma$ whose limit point both lie on the real axis,
define
\be
w_1 = \lim_{t\to-\infty} \gamma(t)\quad\hbox{and}\quad
w_2 = \lim_{t\to\infty} \gamma(t).
\ee
If $w_1 < w_2$, then we can set define $h\in G$ by
\be
h = {1\over \sqrt{w_2-w_1}} \pmatrix{1&-w_1\cr -1& w_2\cr},
\ee
such that $h\cdot\gamma(t)= \xi(t)$.  We make the obvious modifications
to
$h$ if $w_2 < w_1$.  In what follows we will
define the lifting $\tildg$ by taking $\tildg(t) = h\inv\cdot\tildx(t)$,
where $\tildx$ is given in (\ref{tildx}).   From Proposition \ref{RPS}
we immediately obtain the following.

\begin{proposition}\label{Phig}
Suppose $\gamma_0$ is hyperbolic and that $\gamma$ is a
corresponding hypercyclic $\BSko$ curve as in Proposition (\ref{bskhyp}).
Suppose further the limit points $w_1$ and $w_2$ for $\gamma$
lie on the real axis, and that the lifting $\tildg$ is defined as discussed
above.  Then $\ket{\gamma,k}$ is given by the function
\be
\Phi_{\gamma}(z,\zeta) =   A_k B_{k,\tau}
\zeta^k \sum_{g\in\Gamma_0\bs\Gamma} \Bigl({g\cdot z-w_1\over
w_2-g\cdot z}\Bigr)^{ik\tau} {1\over j(g,z)^{2k}}
\biggl[{w_2-w_1\over (w_2 - g\cdot z) (g\cdot z - w_1)}\biggr]^k.
\ee
\end{proposition}

\medskip
Consider the quadratic polynomial
\be\label{wpoly}
{(w_2- z)(z-w_1)\over w_2-w_1},
\ee
which appears on the right side in the preceding proposition.
In \cite{Ka}, Katok associates to a hyperbolic transformation $\gamma_0 =
\pmatrix{a&b\cr c&d\cr}$ the quadratic polynomial
\be
Q_{\gamma_0}(z) = cz^2 + (d-a)z - b.
\ee
Since the roots of this polynomial are the fixed points of $\gamma_0$, $w_1$
and $w_2$, it differs from (\ref{wpoly}) by a constant factor.    In fact
\be
{(w_2- z)(z-w_1)\over w_2-w_1} = - {\sgn \tr (\gamma_0)\over
D_{\gamma_0}^{1/2}}  Q_{\gamma_0}(z),
\ee
where $D_{\gamma_0}$ is the discriminant of $\gamma_0$ as a matrix,  i.e.
$D_{\gamma_0} = (d-a)^2+4bc = (\tr\gamma_0)^2 - 4$.

\medskip\noi
{}From Theorem \ref{rsmnthm} we obtain the following.
\begin{theorem}\label{hypnv}
For $X$ a Riemann surface, the relative Poincar\'e series
\be
\sum_{g\in\Gamma_0\bs \Gamma}
\Bigl({g\cdot z-w_1\over w_2-g\cdot z}\Bigr)^{ik\tau}
{1\over j(g,z)^{2k} Q_{\gamma_0}(g\cdot z)^k}
\ee
associated to a $\BSko$ hypercycle $\gamma$, is non-vanishing for
sufficiently large values of the weight $k$ (such that $k_0$ divides $k$).
\end{theorem}

\bigskip
For the remainder of this subsection we focus on the case of geodesics on
Riemann surfaces.   We begin by illustrating the implications of Theorem
\ref{localuk} to Poincar\'e series.
\begin{theorem}
Suppose $X$ is a Riemann surface and that $\gamma_0\in\Gamma$ is a
dilation.  In a sufficiently narrow band surrounding the imaginary axis
we can estimate
\be\label{xiseries}
\sum_{g \in \Gamma_0\bs\Gamma}  {y^k \over j(g,z)^{2k}}
{1\over (g\cdot z)^k}  \sim e^{ik\left({x\over y} - {\pi\over2}\right)
- {k\over2} {x^2 \over y^2}}
\ee
for $k$ sufficiently large.
\end{theorem}
\begin{proof}
Consider the geodesic $\xi(t) = ie^t$.  The associated state is
\be
\Phi_\xi(z,\zeta) = A_k B_{k,0} \sum_{g\in \Gamma_0\bs\Gamma} {\zeta^k
\over j(g,z)^{2k}} {1\over (g\cdot z)^k},
\ee
where
\be\label{akbko}
A_k B_{k,0} = i^k {2^{2k-2}\over \pi} {2k-2 \choose k-1}\inv,
\ee
A direct application of Theorem \ref{localuk} yields the following:
near the imaginary axis and for $k$ sufficiently large
\be
\Phi_\xi(z,\zeta) \sim \sqrt{{k\over\pi}} e^{ik\left(\theta +
{x\over y}\right) - {k\over2}{x^2\over y^2}}.
\ee
The above result follows because, by Stirling's formula,
\be\label{stirl}
2^{2k-2} {2k-2\choose k-1}\inv \sim \sqrt{\pi k}
\ee
for large $k$.
\end{proof}

\begin{lemma}
Let $r = k^\alpha x$.  If $\alpha > 1/2$ then for fixed $r$ amd $y$,
\be
\left({y\over z}\right)^k \sim e^{ik\left({x\over y} - {\pi\over2}\right)
- {k\over2} {x^2 \over y^2}}
\ee
as $k \to \infty$.
\end{lemma}
This lemma is a simple calculus exercise, which allows us to conclude the
following.
\begin{corollary}
We can find a band surrounding the imaginary axis whose width decreases as
$k^{-\alpha}$ for $\alpha> 1/2$, in which the relative Poincar\'e series
appearing in (\ref{xiseries}) is dominated for large $k$ by the $g =
\hbox{\it id}$ term.
\end{corollary}
Similar results can of course established for the other Poincar\'e series
defined above.

\medskip
We turn next to applications of our results to the relative Poincar\'e
series associated to geodesics by Katok \cite{Ka}.
Given any hyperbolic element
$\gamma_0\in\Gamma$, we have a parametrized geodesic $\gamma(t) =
h\inv\cdot\xi(t)$ with a distinguished lifting $\tildg(t) = h\inv\cdot
\tildx(t)$, where $\tildx$ is given by (\ref{tildx}).   The
resulting states $\ket{\gamma,k}$ can be related to Katok's relative
Poincar\'e  series, $\Theta_{k,\gamma_0}(z)$.  Katok's definition is
\be
\Theta_{k,\gamma_0}(z) := D_{\gamma_0}^{k-1/2} (-\sgn \tr\gamma_0)
{2^{2k-2}\over\pi} {2k-2\choose k-1}\inv
\sum_{g\in\Gamma_0\bs \Gamma}
{1\over j(g,z)^{2k} Q_{\gamma_0}(g\cdot z)^k}.
\ee
Denote the function associated to  $\ket{\gamma,k}$ by
$\Phi_\gamma(z,\zeta)$.  By Proposition \ref{Phig} and (\ref{akbko})
we have
\be
\Phi_\gamma(z,\zeta) = i^k (-\sgn \tr\gamma_0)^{k-1}
D_{\gamma_0}^{-(k-1)/2}\; \zeta^k \Theta_{k,\gamma_0}(z).
\ee

For geodesics there is a nice relation between the intersection
angles $\varphi_p$ and the phases $\omega_p$.   Assume that the fixed
points of $\gamma_0$ are given by real numbers $c_0\pm r_0$.
By the prescription above, we have
\be
\tildg(t) = \Bigl(c_0 + r_0 \tanh\epsilon t + {ir_0\over \cosh t},
\; -r_0 {\sinh \epsilon t + i\over \cosh^2 t}\Bigr),
\ee
where $\epsilon=\pm 1$ depending on the orientation.
We can parametrize the curve by angle instead of arclength by taking $\theta
= \cos\inv(\tanh \epsilon t)$, which gives
\be\label{thparm}
\tildg(\theta) = (c_0 + r_0e^{i\theta}, -r_0 e^{i\theta} \sin\theta),
\ee
where $0<\theta<\pi$.   Now consider the case of two intersecting geodesics,
parametrized by angles $\theta_1$ and $\theta_2$.  Since at the point where
they intersect we have
$r_1 \sin \theta_1 = r_2\sin\theta_2$, it is easy to see from (\ref{thparm})
that the relative phase at such a point is
\be
\omega_p = {\zeta_2\over \zeta_1} = e^{i(\theta_2 - \theta_1)},
\ee
A simple geometric exercise shows that $\theta_2-\theta_1 = \vartheta_p$,
the intersection angle.

\begin{theorem}\label{geothm}
Let $X$ be a Riemann surface and $\gamma_0\in\Gamma$.  Then
$\Theta_{k,\gamma_0}(z)$ is non-vanishing for sufficiently large $k$.
Moreover,
\be
\Vert \Theta_{k,\gamma_0} \Vert_2^2  =
D_{\gamma_0}^{k-1} \left[\left({k\over\pi}\right)^{1/2}T + O(1)\right],
\ee
where $T =  2\cosh\inv({1\over2}\tr\gamma_0)$.
For $\gamma_1,\gamma_2\in \Gamma$ not
conjugate to each other, we have
\begin{eqnarray*}
\braket{\Theta_{k,\gamma_1}}{\Theta_{k,\gamma_2}} & = &
2^{1/2} (D_{\gamma_1}D_{\gamma_2})^{(k-1)/2}
(\sgn \tr\gamma_0 \tr \gamma_1)^{k-1} \\
& & \times \left[ \sum_{p\in [\gamma_1]\cap [\gamma_2]}
{e^{i(k-1/2)\vartheta_p + i\pi/4}\over\sqrt{\sin \vartheta_p}} + O(k^{-
1/2})\right],\\
\end{eqnarray*}
where $[\gamma_j]$ denotes the geodesic on $X$ corresponding to
$\gamma_j$.
\end{theorem}

\medskip
In \cite{Ka}, Katok gives a period formula for the case in which $\Gamma$ is
symmetric (with respect to the imaginary axis), and $\gamma_1$ and
$\gamma_2$ are primitive.  This yields the following exact result for the
imaginary part of the inner product:
\begin{eqnarray*}
\im \braket{\Theta_{k,\gamma_1}}{\Theta_{k,\gamma_2}}
& = & 2^{2k-2}  (D_{\gamma_1}D_{\gamma_2})^{(k-1)/2}
(\sgn \tr\gamma_0 \tr \gamma_1)^{k-1} {2k-2\choose k-1}\inv\\
& &\times \sum_{p\in [\gamma_1]\cap [\gamma_2]}
\sgn(\sin\vartheta_p) P_{k-1}(\cos\vartheta_p),
\end{eqnarray*}
where $P_{k-1}$ is the Legendre polynomial of order $k-1$.  This can be
compared to our formula using (\ref{stirl}) and the theorem of Darboux
on the large $n$ asymptotics of $P_n(\cos\theta)$:
\be
P_{k-1}(\cos\theta) = {\sin((k-1/2)\theta + \pi/4)\over \sqrt{{\pi\over2}
(k-1) \sin\theta}}  + O(k^{-3/2})
\ee
(Theorem 8.21.2 in \cite{Sz}).
The asymptotic estimate of Theorem \ref{geothm} is seen to agree
precisely with Katok's result.

\subsection{Horocycles}

We turn now to the horocycles, curves in $H$ which correspond to parabolic
elements of $\Gamma$.  Note that these do not exist when $X$ is a Riemann
surface, since $\Gamma$ must be hyperbolic in this case.
Horocyclic curves are given either by circles
tangent to the real axis or straight lines parallel to the real axis.
Given a horocycle $\gamma$ and parabolic $\gamma_0\in\Gamma$ such
that
$\gamma_0\cdot\gamma(t) = \gamma(t+T)$, there exists a unique $g\in
G$ such $g\gamma_0 g\inv:z\mapsto z+T$
and $g\cdot\gamma(0)$ lies on the imaginary axis.  The real number
$\lambda = -ig\cdot\gamma(0)$ depends only on $\gamma$ and $T$.
Alternatively, we may use the definition
\be
\lambda = {T\over 2
\sinh\bigl[{1\over2}\rho(\gamma(0),\gamma(T))\bigr]},
\ee
where $\rho$ is the hyperbolic distance.

\begin{proposition}\label{bskhor}
Let $\gamma_0\in\Gamma$ be parabolic, and suppose that $\gamma$ is a
curve $\bbR\to H$ such that $\gamma_0\cdot\gamma(t)=\gamma(t+T)$.
Then $\gamma$ satisfies $\BSk$ if and only if
\be\label{lamt}
\lambda = {kT\over 2\pi m}
\ee
for some $m\in\bbZ$, where $\lambda$ is defined as above.
\end{proposition}\begin{proof}
By equivariance, we assume that $\gamma(t) = i\lambda + z$ and
$\gamma_0:z\mapsto z+T$.
The horizontal lift of $\gamma$ to $G$ is
\be
g(t) = \pmatrix{1 &t\cr 0 &1\cr}
\pmatrix{\lambda^{1/2} &0\cr 0 &\lambda^{-1/2}\cr}
\pmatrix{\cos {t\over 2\lambda}& -\sin {t\over 2\lambda}\cr
\sin {t\over 2\lambda}& \cos {t\over 2\lambda}\cr},
\ee
which corresponds to
\be\label{ltild}
\tildg(t) = (i\lambda+t, \lambda e^{-it/\lambda}).
\ee
The $\BSk$ condition then reduces to the requirement $kT/\lambda = 2\pi
m$ for some $m\in \bbZ$.
\end{proof}

For the remainder of the subsection we assume that $\gamma$ is a $\BSko$
curve and that $\lambda$ and $T$ satisfy
\be
\lambda = {k_0T\over 2\pi m},
\ee
for some $m\in\bbZ$.

\begin{lemma}
Let $\gamma$ be the horocycle $\gamma(t) = i\lambda + t$.  Then
\be
\int_{-\infty}^\infty \psi_{\tildg(t)}(z,\zeta) dt = C_{k,\lambda}\zeta^k
e^{{ikz\over\lambda}},
\ee
where
\be
C_{k,\lambda} = {2\pi(-1)^k\over (2k-1)!}
{k^{2k-1}\over \lambda^{k-1}} e^{-k}
\ee
\end{lemma}
\begin{proof}
The curve $\tildg$ is given by (\ref{ltild}), so that
\be
\int_{-\infty}^\infty \psi_{\tildg(t)}(z,\zeta) dt =
\int_{-\infty}^\infty {\zeta^k \lambda^k e^{{ikt\over\lambda}} \over
(z+i\lambda -
t)^{2k}}
dt.
\ee
The result follows from a contour integration.
\end{proof}

\begin{proposition}
Suppose $\gamma_0\in\Gamma$ is parabolic element that fixes $\infty$
and that $\gamma$ is a corresponding $\BSko$ curve.  The state
$\ket{\gamma,k}$ corresponds to the function
\be
\Phi_\gamma(z,\zeta) = A_k C_{k,\tau}
\sum_{g\in\Gamma/\Gamma_0}
{e^{{ik\over\lambda}g\cdot z}\over j(g,z)^{2k}}.
\ee
\end{proposition}

\begin{corollary}
Let $\Gamma = \SL(2,\bbZ)$, and let $\gamma(t) = {ik\over 2\pi m}+t$,
which
is a $\BSk$ curve.  Then the
state $\ket{\gamma,k}$ is represented by $\zeta^k P_{m,k}(z)$ (up to a
constant depending on $m$ and $k$), where
$P_{m,k}$ is the classical Poincar\'e series:
\be
P_{m,k}(z) = \sum_{g\in\Gamma_{\infty}\bs\Gamma}
{e^{2\pi i m (g\cdot z)}\over j(g,z)^{2k}},
v\ee
with $\Gamma_\infty$ the subgroup fixing $\infty$.
\end{corollary}

\subsection{Circles}

We complete our discussion of specific $\BSk$ curves on $X$ by considering
the circles on $H$.

\begin{proposition}
Let $\gamma$ be a circle in $H$ with (hyperbolic) radius $\mu$, with
$\gamma_0\in\Gamma$  such that $\gamma_0\cdot\gamma(t) =
\gamma(t+T)$  ($\gamma_0$ is either elliptic of finite order or the identity).
Let $n$ denote the minimal integer such that $\gamma_0^n$ is the identity
transformation.  Then $\gamma$ satisfies $\BSk$ if and only if
\be\label{mubsk}
\cosh\mu  =  {nl\over k},
\ee
for some $l\in\bbZ$.
\end{proposition}
\begin{proof}
By equivariance, we assume that the center of the circle is $i$, and that
$\gamma(0) = ie^{-\mu}$.  The location of the center implies that
\be
\gamma_0 = \pmatrix{\cos\pi/n&\sin\pi/n\cr
-\sin\pi/n&\cos\pi/n\cr}.
\ee
Let $a = {\pi\over nT}$.  The curve,
\be
g(t) = \pmatrix{\cos at&\sin at\cr -\sin at&\cos  at\cr}
\pmatrix{e^{-\mu/2}&0\cr 0&e^{\mu/2}\cr}
\pmatrix{\cos bt/2&\sin bt/2\cr \sin bt/2&\cos  bt/2\cr},
\ee
lies over $\gamma$ and will be horizontal provided
\be
b = 2a\cosh\mu.
\ee
The corresponding curve in $SH$ is
\be\label{ctildg}
\tildg(t) = \left({e^\mu\sin at + i\cos at\over e^\mu\cos at - i\sin at}\>,
{e^\mu e^{-ibt}\over (e^\mu\cos at - i\sin at)^2}\right),
\ee
so that $\gamma$ is $\BSk$ iff $kbT = 2\pi l$ for some $l\in\bbZ$, i.e., $b =
{2nl\over k} a$.   The claim follows.
\end{proof}

\medskip\noi
For the remainder of this section, we assume that $\gamma$ is a $\BSko$
curve and that $\mu$ and $l$ satisfy
\be
\cosh \mu = {nl\over k_0},
\ee
for some $l\in\bbZ$.  Note that the $\BSko$ condition requires $nl\ge k_0$.

\begin{proposition}
Let $\gamma$ be a circle in $H$ of radius $\mu$ which is a $\BSk$ curve on
$X$ as above.  Then
\be
\int_0^{nT} \psi_{\tildg(t)}(z,\zeta) = D_{l,k,\mu} F_{nl-k,k}(z,\zeta),
\ee
where $F_{nl-k,k}$ is the element of the orthonormal basis for
$\hilk(H)$ given by (\ref{onbasis}), and
\be
D_{k,\mu} =  2^{1-2k} nT \biggl[ {\pi\over 2k-1} {(nlk/k_0+k-1)!\over
(2k-1)! (nlk/k_0+k)!} \biggr]^{1/2}  {(\sinh \mu/2)^{nlk/k_0-k}
\over (\cosh \mu/2)^{nlk/k_0+k}}
\ee
\end{proposition}

\medskip\noi
{\bf Remark.}  In case $n=1$, the states coming from $\BSk$ circles on
$H$ give an orthonormal basis of $\hilk$.  This is an example of a more
general phenomenon:  {\em If we have a Hamiltonian action of the $n$
torus on the Kahler manifold preserving all structures, our
construction applied to the $\BSk$ level sets of the moment map yield
an orthonormal basis for $\hilk$}.  This actually follows from the
quantum reduction theorem of Guillemin and Sternberg, \cite{GS3}.  In
the case envisioned the reduced spaces are points, and so their
quantization is one-dimensional.  Other examples of this situation
include the Bargmann metric on $\bbC$, where the $\BSk$ circles
correspond to eigenstates of the 1-dimensional harmonic oscillator
problem, and the sphere, where the $\BSk$ circles give spherical
harmonics.

\medskip\noi
\begin{proof}
The curve $\tildg$ is given by (\ref{ctildg}), and we seek to compute
\be
\int_0^{nT} \psi_{\tildg(t)}(z,\zeta) = \int_0^{nT} {\zeta^k e^{k\mu}
e^{ibkt} \over \Bigl[ (e^\mu \cos at + i\sin at) z - (e^\mu\sin at - i\cos
at) \Bigr]^{2k}} \;dt.
\ee
Using the fact that $b = {2nl\over k_0}a$ we can rewrite this as
\be
\int_0^{nT} \psi_{\tildg(t)}(z,\zeta) = 4^k e^{k\mu} \zeta^k  \int_0^{nT}
{e^{2iatk(nl/k_0+1)} \over \Bigl[ (z+i)(e^\mu+1) e^{2iat} + (z-i)(e^\mu-1)
\Bigr]^{2k}} \;dt.
\ee
Changing variables to $u = e^{2iat}$, we have
\be
\int_0^{nT} \psi_{\tildg(t)}(z,\zeta) = 4^k e^{k\mu} \zeta^k \oint
{u^{nlk/k_0+k} \over  \Bigl[ (z+i)(e^\mu+1)u + (z-i)(e^\mu-1)
\Bigr]^{2k}} \;dt,
\ee
where the contour is the unit circle.  Noting that the pole is always
inside the contour, we find
\be
\int_0^{nT} \psi_{\tildg(t)}(z,\zeta) =
nT {nlk/k_0+k-l\choose 2k-1} {(\sinh \mu/2)^{nlk/k_0-k} \over (\cosh
\mu/2)^{nlk/k_0+k}}
\;\zeta^k {(i-z)^{nlk/k_0-k} \over (i+z)^{nlk/k_0+k}}.
\ee
\end{proof}

\begin{proposition}
Suppose $\gamma_0\in\Gamma$ is an elliptic element which fixes $i$, and
that $\gamma$ is a corresponding $\BSko$ circle on $X$.  The state
$\ket{\gamma,k}$ is given by the function
\be
\Phi_{\gamma}(z,\zeta) = A_kD_{k,\mu}
\sum_{g\in\Gamma_0\bs\Gamma_0} F_{nlk/k_0-k, k}(g\cdot(z,\zeta)).
\ee
\end{proposition}

\medskip
If $X$ is a Riemann surface, then the only possibility for $\gamma_0$ is
the identity, so the curves must close on $H$ ($n=1$).

\begin{theorem}
Let $X$ be a Riemann surface.  The relative Poincar\'e series
\be
\sum_{g\in \Gamma} {1\over
j(g,z)^{2k}} {(i-g\cdot z)^{lk/k_0-k} \over (i+g\cdot z)^{lk/k_0+k}},
\ee
is non-vanishing for sufficiently large $k$.
\end{theorem}

\subsection{Towards a geometric construction of a basis}

By Riemann-Roch,
\be\label{5.1}
\mbox{dim } \calS_k\,=\, k(2g-2) - (g-1)\,,
\ee
where $g$ is the genus of $X$.
We now indicate a strategy for choosing, $\forall k$, the same
number of non-intersecting $\BSk$ curves on $X$.  We conjecture that
the associated states form a basis of $\calS_k$.

\medskip
Divide $X$ into $2g-2$ pairs of pants, each bounded by three
simple closed geodesics
(therefore there are $3g-3$ different geodesics on $X$
involved as boundaries).
Consider a pair of pants, $Y$.  By Gauss-Bonnet, it has an area
of $2\pi$.  By the Collar Theorem, there are collar neighborhoods
of the boundary geodesics of $Y$ which are hyperbolic cylinders.
Let $A_c$ denote their total area.  Their complement is the union
of two identical hexagons, let $A_h$ the area of one hexagon
so that $A_c + 2A_h = 2\pi$.

We choose $\BSk$ curves according to the following principle:
\begin{itemize}
\item  On each hyperbolic cylinder, the $\BSk$ hypercycles parallel to the
boundary geodesic.
\item  In each of the hexagons, the $\BSk$ curves of a function with a
single critical point in the interior.
\end{itemize}
Now count $\BSk$ curves in each region:
\begin{proposition}
The above scheme produces exactly $\mbox{dim }\calS_k$ $\BSk$ curves,
$\forall k$.
\end{proposition}


\end{document}